\documentclass[%
reprint,
aps,
superscriptaddress
]{revtex4-2}

\usepackage{times}
\usepackage{amsmath}
\usepackage{amssymb}
\usepackage{graphicx}
\usepackage{bm}
\usepackage[alsoload=synchem, per-mode=symbol]{siunitx}
    \DeclareSIUnit \ev{eV}
    \DeclareSIUnit \sccm{sccm}
\usepackage{braket}
\usepackage{xspace}
\usepackage{float}
\usepackage{changepage} 
\usepackage{enumitem}
\usepackage{tabularx,colortbl} 
\usepackage{multirow}
\usepackage{afterpage}
\usepackage{color}
\definecolor{lightGray}{rgb}{0.863, 0.863, 0.863}
\definecolor{lightOrange}{rgb}{0.933, 0.553, 0.388}

\newcolumntype{Y}{>{\centering\arraybackslash}X} 
\newcommand{\RN}[1]{\textup{\uppercase\expandafter{\romannumeral#1}}} 

\def\g2{g^{(2)}(t)}
\newcommand{\autocorr}[1]{g^{(2)}(#1)}

\begin{document} 

\preprint{APS/123-QED}

\title{Probing the Optical Dynamics of Quantum Emitters in Hexagonal Boron Nitride}

\author{Raj N. Patel}
\affiliation{
Quantum Engineering Laboratory, Department of Electrical and Systems Engineering, University of Pennsylvania, Philadelphia, PA 19104, United States
}

\author{David A. Hopper}
\thanks{
Present address: MITRE Corporation, 7515 Colshire Dr.
McLean, VA 22102, USA
}
\affiliation{ 
Quantum Engineering Laboratory, Department of Electrical and Systems Engineering, University of Pennsylvania, Philadelphia, PA 19104, United States
}

\author{Jordan A. Gusdorff}
\affiliation{
Quantum Engineering Laboratory, Department of Electrical and Systems Engineering, University of Pennsylvania, Philadelphia, PA 19104, United States
}
\affiliation{
Department of Materials Science and Engineering, University of Pennsylvania, Philadelphia, PA 19104, USA
}

\author{Mark E. Turiansky}
\affiliation{
Department of Physics, University of California, Santa Barbara, CA 93106, United States
}

\author{Tzu-Yung Huang}
\affiliation{ 
Quantum Engineering Laboratory, Department of Electrical and Systems Engineering, University of Pennsylvania, Philadelphia, PA 19104, United States
}

\author{Rebecca E. K. Fishman}
\affiliation{ 
Quantum Engineering Laboratory, Department of Electrical and Systems Engineering, University of Pennsylvania, Philadelphia, PA 19104, United States
}
\affiliation{
Department of Physics and Astronomy, University of Pennsylvania, Philadelphia, PA 19104, USA
}

\author{Benjamin Porat}
\thanks{
Present address: Raytheon Intelligence and Space, 2000 E El Segundo Blvd, El Segundo, CA 90245
}\affiliation{
Quantum Engineering Laboratory, Department of Electrical and Systems Engineering, University of Pennsylvania, Philadelphia, PA 19104, United States
}

\author{Chris G. Van de Walle}
\affiliation{
Materials Department, University of California, Santa Barbara, CA 93106, United States
}

\author{Lee C. Bassett}
\email[Corresponding author.  Email: ]{lbassett@seas.upenn.edu}
\affiliation{ 
Quantum Engineering Laboratory, Department of Electrical and Systems Engineering, University of Pennsylvania, Philadelphia, PA 19104, United States
}

\date{\today}

\begin{abstract}
Hexagonal boron nitride is a van der Waals material that hosts visible-wavelength quantum emitters at room temperature. 
However, experimental identification of the quantum emitters’ electronic structure is lacking, and key details of their charge and spin properties remain unknown. 
Here, we probe the optical dynamics of quantum emitters in hexagonal boron nitride using photon emission correlation spectroscopy. 
Several quantum emitters exhibit ideal single-photon emission with noise-limited photon antibunching, $\bm{g^{(2)}{(0)}=0}$.
The photoluminescence emission lineshapes are consistent with individual vibronic transitions. 
However, polarization-resolved excitation and emission suggests the role of multiple optical transitions, and photon emission correlation spectroscopy reveals complicated optical dynamics associated with excitation and relaxation through multiple electronic excited states.
We compare the experimental results to quantitative optical dynamics simulations, develop electronic structure models that are consistent with the observations, and discuss the results in the context of \textit{ab initio} theoretical calculations.
\end{abstract}

\maketitle


\section{Introduction}

Hexagonal boron nitride (h-BN) is a wide-bandgap ($\sim$6 eV) van der Waals material that hosts fluorescent quantum emitters (QEs) at room temperature \cite{Meuret2015,Tran2016a,Chejanovsky2016,Martinez2016a,Shotan2016,Exarhos2017,Noh2018}. 
The QEs in h-BN are bright and photostable with narrow emission linewidths and high single-photon purity, as required for quantum technologies \cite{Exarhos2017,Chakraborty2019,Kim2019d,Dietrich2020}.
Recent observations of room-temperature magnetic field dependence and spin resonance of QEs in h-BN make them attractive for spin-based quantum sensing and computation \cite{Exarhos2019,Gottscholl2020,Chejanovsky2021,Stern2021,Gottscholl2021}.

Despite intense interest in h-BN's QEs, their chemical and electronic structures remain uncertain, as do key details regarding their optical, spin, and charge dynamics.
The pronounced heterogeneity of observations suggests that QEs originate from multiple distinct defect structures \cite{Tran2016,Bourrellier2016,Exarhos2017,Ziegler2018,Stern2019}.
Ultraviolet emission around 4.1~eV has been attributed to the carbon dimer $C_B C_N$ \cite{Mackoit-Sinkeviciene2019}, whereas near-infrared emission around 1.7~eV and an associated optically detected magnetic resonance signal is attributed to the negatively-charged boron vacancy, $V_B^-$ \cite{Gottscholl2020}. 
For QEs in the visible spectrum, experiments utilizing various forms of electron and optical microscopy, spectroscopy, and materials growth and treatments have generated a detailed, yet complicated, empirical understanding of the QEs' creation, stabilization, and principal optical signatures \cite{Hayee2020,Cretu2021Atomic-ScaleDefects,Feng2018ImagingResolution,Chejanovsky2016,Hernandez-Minguez2018,Hou2018a,Boll2020,Jungwirth2016,Jungwirth2017,Toledo2018a,Comtet2019a,Grosso2020,NgocMyDuong2018,My2019,Mendelson2020,Breitweiser2020,Fournier2021,Xu2021CreatingSubstrates,Stewart2021QuantumNitride,Glushkov2021DirectCharacterization}.
Theoretical work suggests that vacancies and their complexes, along with substitutional carbon atoms and dangling bonds, are likely candidates, although consensus is still lacking \cite{Tawfik2017,Weston2018,Sajid2018,Turiansky2019a,Korona2019a,Dev2020}. 
Specific candidates include $V_{\rm N}N_{\rm B}$, $V_{\rm N}C_{\rm B}$, $V_{\rm B}C_{\rm N}$, and the boron dangling bond.

Even less is known about the visible QEs' optical dynamics.
Optical dynamics arise from a QE's electronic structure together with radiative and nonradiative transitions between electronic states.
State transitions can involve multiple processes including electron-phonon interactions, intersystem crossings between different spin manifolds, and ionization or recombination events. 
For QEs in h-BN, previous studies have reported photon bunching associated with metastable dark states \cite{Chejanovsky2016,Exarhos2017,Exarhos2019}, and yet the nature of these states and the transitions between them remains unclear.
Some QEs exhibit magnetic-field-dependent modulation of their photoluminescence (PL) signal, consistent with a spin-dependent intersystem crossing, whereas others do not  \cite{Exarhos2019,Stern2021}.
An optically detected magnetic resonance signal was observed for a particular QE under excitation at 633~nm but not at 532 nm \cite{Chejanovsky2021}.
Such observations present a complicated picture of the visible QEs, likely involving multiple defect classes (\textit{e.g.}, different chemical structures or charge states), strong local perturbations, and complex excitation and relaxation pathways.
Improved understanding of the QEs' optical dynamics can resolve these mysteries.
Such understanding is also a prerequisite to designing quantum control protocols that would facilitate their use in quantum technologies.
Here, we use quantitative spectral, spatial, and temporal PL spectroscopy to investigate the optical dynamics of h-BN's QEs.

Photons emitted by a QE carry a wealth of information about its electronic structure and optical dynamics. 
For vibronic optical transitions, the photon energy and polarization distributions reflect the details of electron-phonon coupling and optical dipole selection rules, respectively.
The QEs in h-BN generally exhibit linearly-polarized PL and strong electron-phonon coupling associated with a single vibronic transition \cite{Exarhos2017}, and yet other experimental and theoretical evidence points to the involvement of multiple excited states in the optical dynamics \cite{Jungwirth2017,Turiansky2019a,Dietrich2020,Sajid2020a}.
Time-dependent measurements provide complementary information.
The second-order photon autocorrelation function, $\autocorr{\tau}$, is widely used to identify single-photon emitters. 
As a more general analytical tool, photon emission correlation spectroscopy (PECS) yields quantitative information about a QE's optical dynamics \cite{Fishman2021}.
Qualitatively, we distinguish between photon antibunching ($\autocorr{\tau}<1$) as a signature of non-classical light, with single-photon emission as a special case when $\autocorr{0}=0$, and photon bunching ($\autocorr{\tau}>1$ for $\tau\neq0$) as a signature of dark, metastable states accessed via nonradiative transitions.
Quantitative measurements of $\autocorr{\tau}$ as a function of optical excitation power or wavelength can elucidate a QE's excitation and emission pathways as well as bunching mechanisms.

Prior observations of h-BN's visible QEs feature both bunching and antibunching signatures, although with some unusual, conflicting patterns.  
Some QEs respond to applied dc and ac magnetic fields in a manner consistent with spin-mediated intersystem crossing transitions, whereas others do not \cite{Exarhos2019,Chejanovsky2021,Stern2021}.
A lack of evidence for pure single-photon emission motivated a proposal that h-BN's QEs occur in pairs as ``double defects''  \cite{Bommer2019a}.
In this work, we compare quantitative PL spectroscopy and PECS measurements of h-BN's QEs with theoretical simulations.
We show that QEs in room-temperature h-BN can exhibit pure single-photon emission, with $\autocorr{0}=0$ within experimental uncertainty.
Furthermore, we find evidence for multiple electronic states connected by radiative and nonradiative transitions, with associated timescales spanning over five orders of magnitude.
Comparing the experiments to theoretical proposals, we find that the boron dangling bond model provides a consistent, quantitative understanding of the observations for individual QEs as well as their heterogeneity.

\section{Results}

The Results and Discussion are organized as follows.
First, we report the basic optical characteristics of five well-isolated QEs across three samples, specifically including their PL spectra, PL saturation as a function of excitation power, polarization properties, and single-photon purity. 
Next, we investigate the QEs' optical dynamics using PECS as a function of excitation power and wavelength.
We consider different models for the electronic level structure and simulate the corresponding optical dynamics.
Finally, we compare the theoretical simulations to experimental observations and discuss the implications for understanding the QEs' electronic and chemical structure along with their spin and charge dynamics.

\subsection{Photoluminescence characterization}

We used a custom-built confocal microscope to study individual QEs in h-BN under ambient conditions. 
The h-BN bulk crystals were mechanically exfoliated into thin flakes using a dry transfer process \cite{Huang2015} and transferred to a SiO$_{2}$/Si substrate patterned with circular trenches \cite{Exarhos2017}.
Prior to the optical studies, the samples were annealed in a tube furnace at \SI{850}{\celsius} in Ar atmosphere for 2 hours. 
This annealing process has been shown to brighten the emitters \cite{Breitweiser2020}.
The QEs are illuminated with either of two continuous-wave lasers operating at 532 nm and 592 nm wavelengths, where excitation power and polarization are controlled.
To differentiate between excitation wavelengths in this work, data recorded under 532 nm (592 nm) excitation are plotted in green (orange) in the relevant figures.
Some QEs disappeared during experiments, hence the set of measurements is not identical for each QE.
See Materials and Methods for additional details on sample preparation, data acquisition, and analysis.

\begin{figure*}[p]
    \centering
    \includegraphics[width=7in]{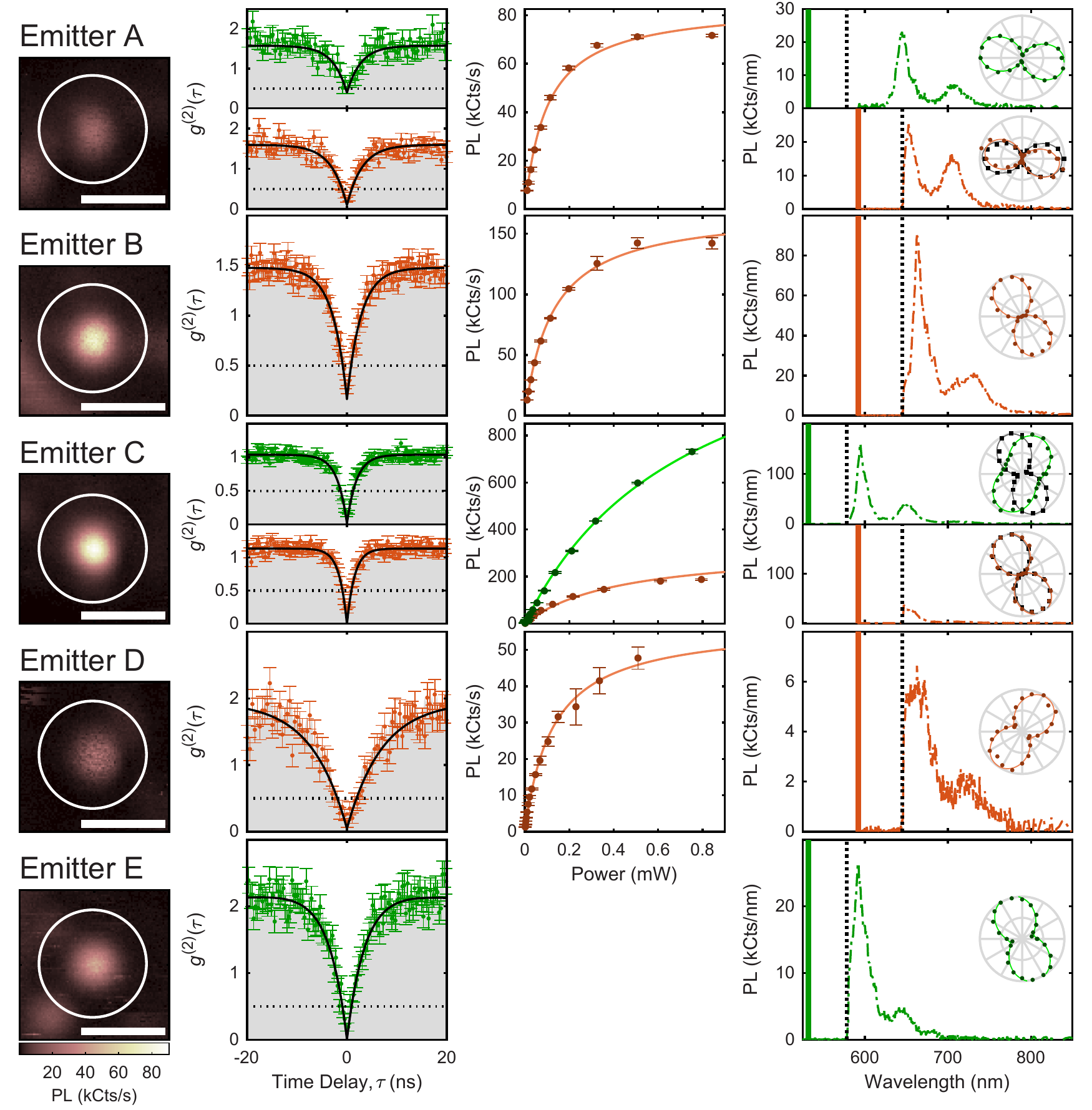}
    \caption{
    \textbf{Photoluminescence characterization.}  
    In all panels, data plotted in green (orange) were acquired under 532~nm (592~nm) excitation. 
    \textbf{(Column 1)} $\micro$-PL images of the QEs (circled), acquired under 592~nm (QEs A-D) or 532~nm (QE E) excitation.  Scale bars denote 1 $\micro$m.
    \textbf{(Column 2)} Second-order photon autocorrelation function (colored points), fit using an empirical model discussed in the text (black curve).  Error bars represent Poissonian uncertainties based on the photon counts in each bin. 
    \textbf{(Column 3)} Steady-state, background-subtracted PL intensity as a function of excitation power (points), fit using an empirical saturation model discussed in the text (solid curves). Error bars represent one standard deviation based on three measurement repeats.
    \textbf{(Column 4)} PL spectra and polarization data. Vertical colored lines represent the excitation laser wavelengths, and black dotted lines indicate cut-on wavelengths for long-pass optical filters in the collection path. 
    Insets: PL intensity as a function of linear excitation polarization angle (colored circles) or filtered by linear polarization angle in emission (black squares). Solid curves are fits to the data using an empirical model discussed in the text.
    }
    \label{fig:summary}
\end{figure*}

Figure \ref{fig:summary} summarizes the PL characterization measurements.
Each row corresponds to a particular QE (labeled A-E), and each column corresponds to a different experiment.
The first column includes $\micro$-PL images of each QE, acquired by scanning a fast steering mirror and recording the accumulated counts at each pixel.
A two-dimensional Gaussian fit to each $\micro$-PL image yields the background and signal levels for subsequent studies.
The second column displays $\autocorr{\tau}$ measurements over short delay times, showing characteristic antibunching dips fit by an empirical model for a multi-level system (see Materials and Methods).
The third column displays the steady-state PL signal as a function of excitation power.
These data are fit using an empirical saturation model,
\begin{align}
    C(P) = \frac{C_{\mathrm{sat}}^{\lambda}P}{P + P_{\mathrm{sat}}^{\lambda}}
    \label{eq:saturationCurve}
\end{align}
where 
$C$ is the background-subtracted, steady-state PL count rate,
$P$ is the optical excitation power,
$C_{\mathrm{sat}}^{\lambda}$ is the saturation count rate for a particular excitation wavelength, $\lambda$, 
and $P_{\mathrm{sat}}^{\lambda}$ is the corresponding saturation power.
The best-fit results are reported in Supplementary Table S2.

The fourth column of Fig.~\ref{fig:summary} presents PL emission spectra and polarization measurements.
In each PL spectrum, the long pass filter cut-on wavelength is indicated as a vertical dotted line, and the excitation wavelength is a solid line.
The inset to each PL spectra panel presents measurements of the QE's excitation and emission polarization properties.
These data are acquired by varying the linear polarization of the excitation laser (colored circles) or by passing the PL through a linear polarizer placed in the collection path (black squares).
For excitation polarization measurements, the linear polarizer in the collection path is removed.
For emission polarization measurement, the excitation polarization is set to maximize the PL.
At each polarization setting, we record the steady-state PL intensity as well as a background intensity from a spatial location offset $\SI{\sim1}{\micro\meter}$ from the QE, which is subtracted to yield the PL signal.
The order of the polarization angles is set randomly to minimize effects of drift and hysteresis.
Solid curves are fits to the data using the model function
\begin{equation}
    I_{s}^{\lambda}(\theta) = A_{s}^{\lambda} \cos^{2}(\theta-\theta_{s}^{\lambda}) + B_{s}^{\lambda} \label{eqn:dipole}
\end{equation}
where $\lambda$ indicates the excitation wavelength, $s$ indicates excitation (ex) or emission (em), $A_{s}^{\lambda}$ is the amplitude, $\theta_{s}^{\lambda}$ is the polarization angle of maximum intensity, and $B_{s}^{\lambda}$ is the offset.
From the fit results, the visibility is calculated as 
\begin{equation}
    V_{s}^{\lambda} = \frac{I^\mathrm{max}_s-I^\mathrm{min}_s}{I^\mathrm{max}_s+I^\mathrm{min}_s} = \frac{A_{s}^{\lambda}}{A_{s}^{\lambda} + 2B_{s}^{\lambda}} \label{eqn:visibility}    
\end{equation}
where $I^{\mathrm{max}}_s$ and $I^{\mathrm{min}}_s$ are the maximum and minimum intensities, respectively.
The misalignment between the excitation and emission polarization angles is
\begin{equation}
\Delta\theta^{\lambda} = \theta_{\mathrm{ex}}^{\lambda} - \theta_{\mathrm{em}}^{\lambda}
\label{eq:polarization_alignment}
\end{equation}
The best-fit parameters are reported in Supplementary Table S2.

\subsection{Spectral emission lineshapes}

\begin{figure}[b]
  \centering
  \includegraphics[width=3.375in]{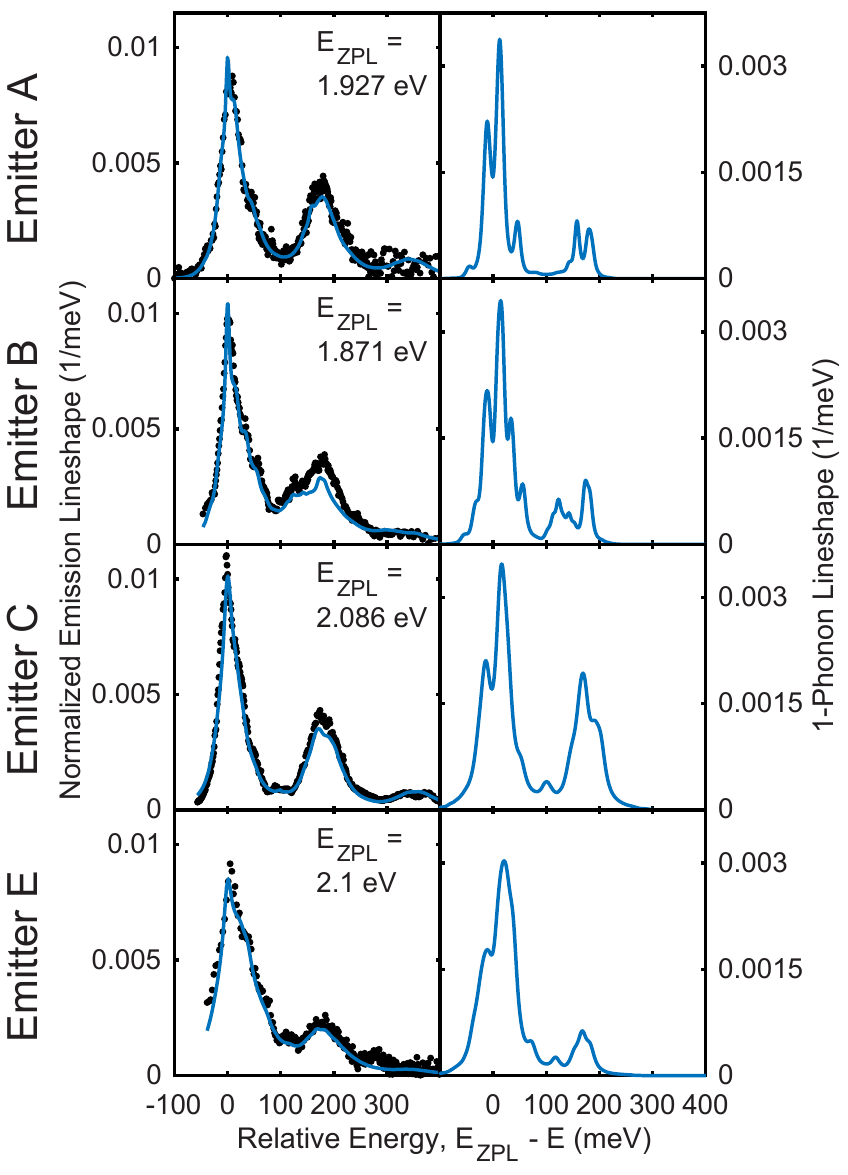}
  \caption{\textbf{PL emission lineshapes.}
  \textbf{(Left Column)} Observed spectral emission lineshapes (black points) and fits according to Huang-Rhys theory (blue curve). Experimental uncertainties are comparable to the size of the data points.
  \textbf{(Right Column)} One-phonon vibronic coupling lineshape corresponding to the fits at left.
  }
  \label{fig:lineshapes}
\end{figure}

For spectra that are not cut off by the excitation filter (namely, the 532~nm excitation spectra for QEs A, C, and E, and the 592~nm spectrum for QE B), we find that the lineshapes are consistent with a Huang-Rhys model for a vibronic transition associated with a single zero-phonon line (ZPL).
We use the analysis method described by Ref.~\cite{Exarhos2017} to fit the observed PL spectra using an empirical model in which the ZPL energy, ZPL width, Huang-Rhys factor, and vibronic coupling lineshape are free parameters; see Materials and Methods for additional details.
The results are shown in Fig.~\ref{fig:lineshapes}.
In the left column, we plot the normalized observed emission lineshape, $L(\Delta E) \propto S(\Delta E)/E^3$, where $S(\Delta E)$ is the spectral intensity distribution as a function of the relative energy $\Delta E=E_\mathrm{ZPL}-E$, with $E$ denoting the photon emission energy and $E_\mathrm{ZPL}$ denoting the ZPL energy.
The factor $1/E^3$ accounts for the photon energy dependence in spontaneous emission.
Each solid curve is the result of a weighted least-squares fit of the model to the experimental lineshapes.
The right column of Fig.~\ref{fig:lineshapes} shows the corresponding 1-phonon vibronic coupling lineshape for each fit.
Best-fit parameters are reported in Supplementary Table S2.

\subsection{Photon emission correlation spectroscopy}

Temporal correlations between fluorescence photons reveal information about a QE's excitation and emission dynamics.
In this work, we use PECS for two purposes: to verify the single-photon purity of the QEs and to probe their optical dynamics as a function of optical excitation rate.
We calculate $\autocorr{\tau}$ from the photon arrival times acquired from two detectors in a Hanbury Brown and Twiss interferometer using a time-correlated single-photon counting module.
For QEs in h-BN, the timescales over which antibunching and bunching occur can vary over at least 6 orders of magnitude \cite{Exarhos2017,Exarhos2019,Tran2016}.
For this reason, we initially calculate and analyze $\autocorr{\tau}$ over a logarithmic scale spanning from 100~ps to 1~s.
We fit the background-corrected data using a general empirical model for a QE's optical dynamics with a varying number of levels:
\begin{equation}
    \autocorr{\tau} = 1 - C_{1}e^{-\gamma_{1}|\tau|} + \sum_{i=2}^{n}C_{i}e^{-\gamma_{i}|\tau|}
    \label{eq:generalizedG2}
\end{equation}
Here, $\gamma_{1}$ is the antibunching rate, $C_1$ is the antibunching amplitude, $\gamma_{i}$ for $i\ge 2$ are bunching rates, and $C_i$ for $i\ge 2$ are the corresponding bunching amplitudes.
We determine the number of resolvable timescales, $n$, by calculating and comparing the Akaike Information Criterion (AIC) and the reduced chi-squared statistic for each best-fit model.
In optical dynamics models, an $N$-level system is characterized by $N-1$ rates, corresponding to the nonzero eigenvalues of the generator matrix (see, \textit{e.g.}, Eq.~(\ref{eq:generatorMatrix}) in Materials and Methods).
Therefore, the inferred value of $n$ places a lower limit on the number of electronic levels required to describe the observations, $N\geq n+1$.
We extract the rates, amplitudes, and their corresponding uncertainty from these fits for comparisons with theoretical simulations.
In order to assess the single-photon purity associated with the value of $\autocorr{0}$, we perform a subsequent analysis of $\autocorr{\tau}$ calculated over a linear scale of delay times, $\tau\in [-20,20]$~ns.
Examples of such data are shown in Fig.~\ref{fig:summary}, along with constrained fits in which only the antibunching parameters $\gamma_1$ and $C_1$ are allowed to vary, and which account for the instrument response function associated with detector timing jitter.
See Materials and Methods for further details.

\subsection{Verifying single-photon emission}


\begin{figure}[b!]
  \centering    
  \includegraphics[width=3in]{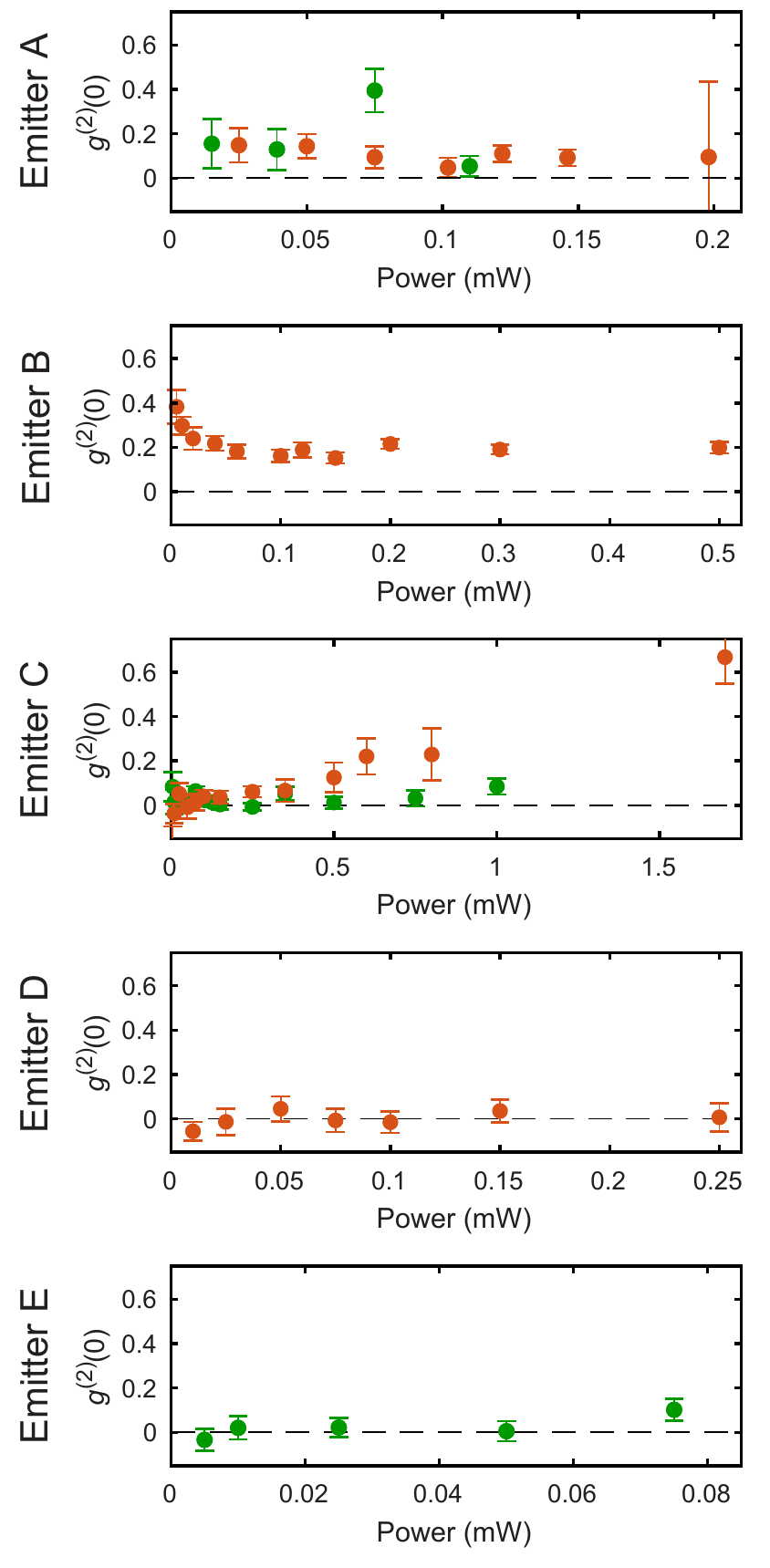}
  \caption{
  \textbf{Single-photon emission characteristics.}
  Zero-delay photon autocorrelation function as a function of optical excitation power. Orange (green) data correspond to excitation at 592~nm (532~nm). 
  Error bars represent one standard deviation derived from fits as described in the text.
  }
  \label{fig:irfg2}
\end{figure}

Any observation of sub-Poissonian statistics, $\autocorr{0}<1$, indicates the presence of quantized photon emission.
The threshold $\autocorr{0}<0.5$ is often used to indicate single-photon emission; however, a more precise interpretation is that a PL signal is dominated by a single-photon emitter when $\autocorr{0}<0.5$ \cite{Fishman2021}.
An observation of $\autocorr{0}>0$ implies a non-zero probability of observing two detection events simultaneously, either due to background fluorescence, detection timing jitter, or the presence of multiple QEs.
Studies of h-BN's QEs routinely report $\autocorr{0}<0.5$, however we are unaware of any prior room-temperature observations of pure single-photon emission with $\autocorr{0}=0$.
Partially on the basis of such observations, Ref.~\cite{Bommer2019a} proposed that h-BN's QEs actually occur in pairs as double defects with parallel emission pathways.

We find that QEs in h-BN can indeed exhibit pure single-photon emission at room temperature.
Figure \ref{fig:irfg2} shows $g^{(2)}(0)$ for each QE as a function of excitation power.
These data are corrected for background fluorescence and detector timing jitter, as described in Materials and Methods.
For QEs C, D and E, we observe $\autocorr{0}=0$ within the experimental uncertainty, particularly at low excitation powers. 
For QEs A and B, we observe $\autocorr{0}\sim0.1-0.2$.
The offset from zero could reflect a contribution from additional dim emitters, however we believe it is more likely to result from incomplete estimation of the background. 
For instance, QE B sits on an extended background feature whose contribution is not captured by our standard analysis method. 
For QE C, we attribute the increase in $\autocorr{0}$ as a function of excitation power to limitations in the instrument-response-function correction as the antibunching rate exceeds the detector timing resolution.

\begin{figure*}[htb!]
    \centering
    \includegraphics[width=7in]{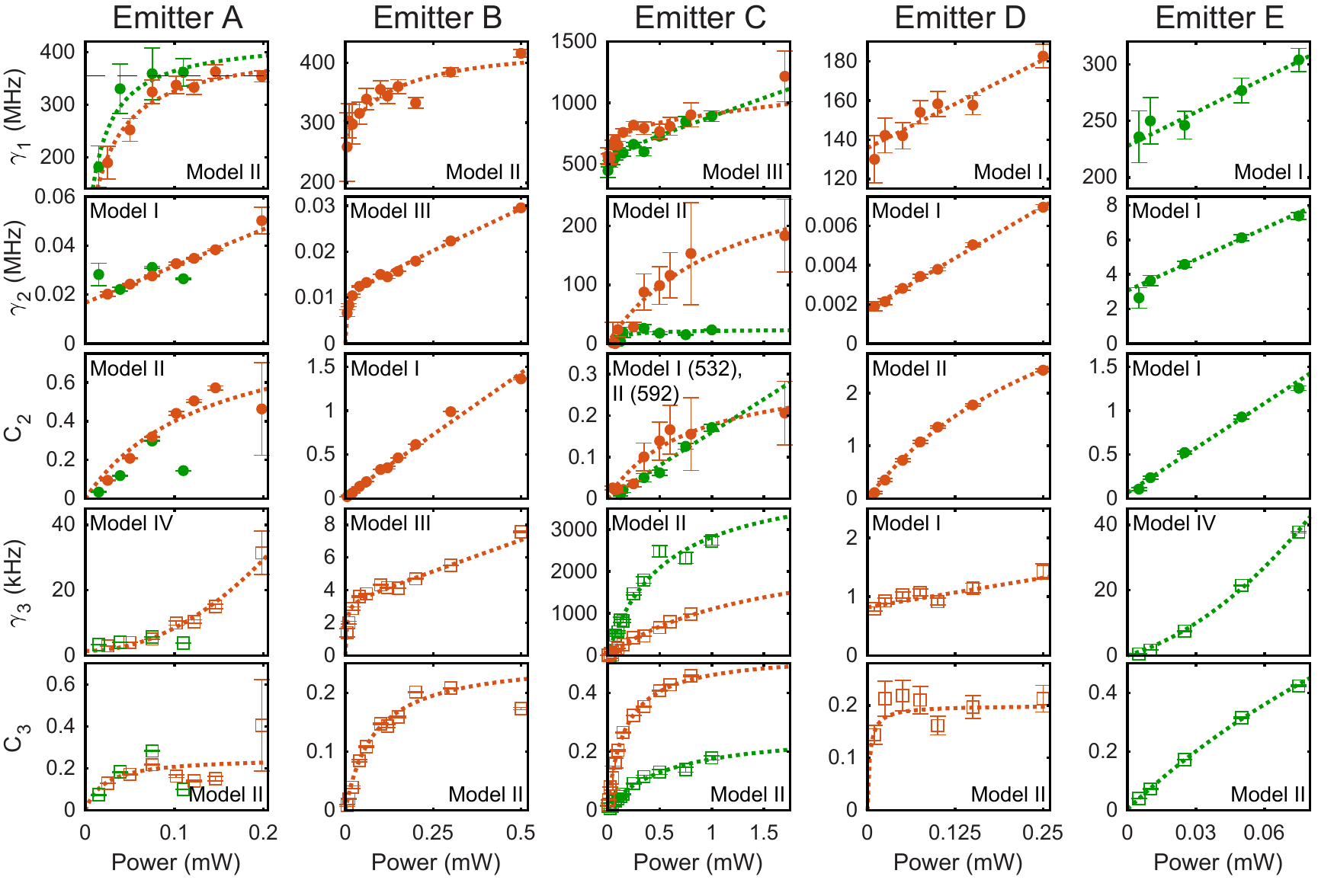}
    \caption{\textbf{Photon emission correlation spectroscopy.}
    Excitation power and wavelength dependence of \textbf{(Row 1)} the antibunching rate ($\gamma_{1}$) denoted as circles, \textbf{(Row 2)} the bunching rate ($\gamma_{2}$) denoted as circles, \textbf{(Row 3)} the bunching amplitude ($C_{2}$) denoted as circles, \textbf{(Row 4)} the bunching rate ($\gamma_{3}$) denoted as squares and \textbf{(Row 5)} the bunching amplitude ($C_{3}$) denoted as squares, of the five QEs presented in each column.
    The black dashed line (QE A) represents the lifetime.
    Orange (green) data correspond to excitation at 592~nm (532~nm).
    The error bars represent one standard deviation.
    The dotted lines are fits to the metadata as discussed in the text.}
    \label{fig:photonStats}
\end{figure*}

\subsection{Probing the optical dynamics}

Figure \ref{fig:photonStats} summarizes the results of fitting the empirical model of Eq.~(\ref{eq:generalizedG2}) to PECS measurements as a function of optical excitation power.
The figure includes the best-fit antibunching rate (top row) as well as the first two bunching rates and amplitudes (lower rows).
The PECS data for QEs B, D, and E are best described by a three-timescale  model ($n=3$), whereas QE A exhibits four resolvable timescales ($n=4$). 
For QE C, we resolve two or three timescales depending on the excitation power and wavelength.
The best-fit results for QE A's third bunching component ($\gamma_4$ and $C_4$), as well as the antibunching amplitude ($C_1$) for all emitters are shown in Supplementary Figures S6 and S7, respectively. 
As in previous figures, colors indicate the optical excitation wavelength.
The PL decay rate of QE A was directly measured to be \SI{355}{\mega\hertz} using a picosecond pulsed laser (see Section S6 and Fig. S4 in the Supplementary Materials); this measurement is shown in Fig.~\ref{fig:photonStats}(a) as a dashed black line.
The PL lifetime measurement was only performed for QE A given the susceptibility of h-BN's QEs to disappear under pulsed excitation.

To fit these metadata, we consider the following empirical models:
\begin{subequations}
\begin{align}
    &\text{Model I (Linear)}: \nonumber \\ 
    &\qquad  R(P) = R_{0}+m_{0}P \label{eq:linearFuncForm} \\
    &\text{Model II (First-Order Saturation)}: \nonumber \\
    &\qquad  R(P) = R_{0}+\frac{R_{sat}P}{P+P_{sat}}
    \label{eq:saturationFuncForm}\\
    &\text{Model III (Second-Order Saturation)}: \nonumber \\
    &\qquad  R(P) = R_{0}+\frac{(m_{0}P_{sat}P+m_{1}P^2)}{P+P_{sat}}
    \label{eq:quadraticSaturationFuncForm} \\
    &\text{Model IV (Quadratic)}: \nonumber \\
    &\qquad  R(P) = R_{0}+m_{0}P+m_{1}P^2
    \label{eq:quadraticFuncForm}
\end{align}
\end{subequations}
where $P$ is the excitation power and the other variables are free parameters representing zero-power offset, $R_{0}$, low-power slope, $m_{0}$, high power slope, $m_{1}$, saturation value, $R_{sat}$, and saturation power, $P_{sat}$.
Dotted curves in Fig.~\ref{fig:photonStats} show the best-fit results for the model listed in each panel; 
in each case, we select the model with the fewest free parameters that qualitatively fits the data.
Best-fit parameters and uncertainties for each fit are reported in Supplementary Tables S5 and S6.

The antibunching rate, $\gamma_{1}$, exhibits a markedly nonlinear power dependence for QEs A, B, and C whereas the dependence appears to be linear for QEs D and E.
However, we note that the power range in the data for QEs D and E might not be large enough for nonlinearities to emerge.
For comparison, QEs B and C are excited with up to $\sim 4P_\mathrm{sat}^\lambda$ whereas QE D is excited with up to $\sim 2P_\mathrm{sat}^\lambda$ (see Supplementary Table S2).
The zero-power antibunching-rate offset ($R_0$) for QEs B-E is clearly nonzero, whereas the fits using Model II for QE A are poorly constrained, yielding $R_0=0\pm261$ MHz and $R_0=0\pm167$ MHz for green and orange excitation, respectively.
The antibunching amplitudes (see Supplementary Figure S7) for all QEs show a nonlinear saturation dependence on excitation power with an expected convergence to  $C_1\sim1$ at zero excitation power.

The bunching dynamics exhibit significant quantitative and qualitative variations across emitters.
The fastest bunching rate, $\gamma_2$, scales linearly with excitation power and has a non-zero offset for QEs A, D and E, whereas it exhibits saturation behavior and zero offset for QEs B and C.
The magnitudes of $\gamma_2$ range from several kilohertz (QEs A, B, and D) up to several megahertz or faster (QEs C and E).
The slower bunching rate, $\gamma_3$, exhibits the largest qualitative variation across emitters, including linear (QE D), quadratic (QEs A and E), and saturation models (QEs B and C).
Only QE D exhibits clear evidence for a non-zero offset for $\gamma_3$.
The magnitudes of $\gamma_3$ are typically in the kilohertz range, with the exception of QE C, whose $\gamma_3$ increases beyond 1~MHz at high powers. 
The bunching amplitudes primarily depend nonlinearly on excitation power, except for $C_{2}$ of QEs B, C (green excitation) and E, which scale linearly with excitation power. 
All of the bunching-amplitude fits are consistent with zero offset, except for QE E, where small residual offsets ($R_0<0.1$) likely reflect minor systematic errors in the analysis or inaccuracies of the empirical models.
For QE A, we restrict the meta-analysis of bunching parameters to the orange-excitation data, which extend to higher excitation power.
However, we note that the green-excitation bunching parameters generally track the data for orange excitation.

\subsection{Electronic model and optical dynamics simulations}

\begin{figure*}[t!]
  \centering
  \includegraphics[width=6.75in]{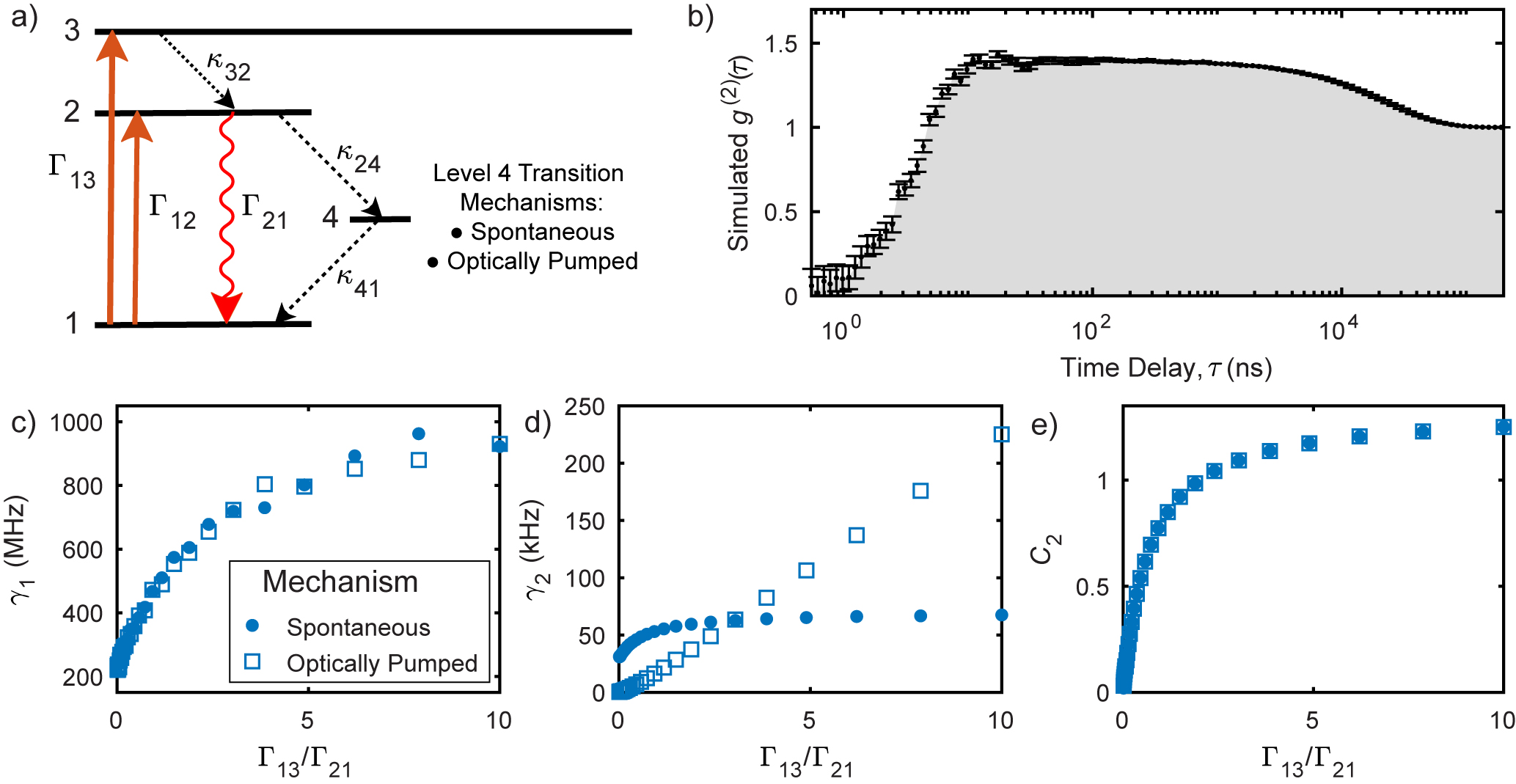}
  \caption{\textbf{Electronic level structure and optical dynamics simulations.}
  \textbf{(a)} An investigative four-level electronic model consisting of the ground state (level 1), radiative state (level 2), excited state (level 3), and metastable state (level 4).
  Orange arrows represent excitation pathways (with rates $\Gamma_{12}$ and $\Gamma_{13})$, the wavy red arrow represents radiative emission (with rate $\Gamma_{21}$), and dotted black arrows represent nonradiative transitions (with rates $\kappa_{32}$, $\kappa_{24}$ and $\kappa_{41}$). 
  \textbf{(b)} Simulated $\autocorr{\tau}$ for $\Gamma_{12}=0$, $\Gamma_{13}=84$ MHz, $\Gamma_{21}=300$ MHz, $\kappa_{32}=600$ MHz, $\kappa_{24}=60$ kHz, and $\kappa_{41}=30$ kHz. Error bars represent simulated photon shot noise.
  \textbf{(c-e)} Best-fit parameters $\gamma_1$, $\gamma_2$, and $C_2$ determined by fitting simulated $\autocorr{\tau}$ data using Eq.~\ref{eq:generalizedG2} with $n=2$.
  The results are plotted as a function of $\Gamma_{13}/\Gamma_{21}$, where $\Gamma_{21}=300$ MHz is a fixed parameter.
  }
  \label{fig:simulations}
\end{figure*}

We find that the key features observed in Fig.~\ref{fig:photonStats} can be understood using the four-level electronic model shown in Fig.~\ref{fig:simulations}(a).
Figure~\ref{fig:simulations} summarizes the results of optical dynamics simulations for this model.
Given a set of transition rates for the model, we simulate $\autocorr{\tau}$ including the effects of timing resolution and shot noise (e.g., Fig.~\ref{fig:simulations}(b)), and we subsequently fit the simulated data using the empirical model of Eq.~(\ref{eq:generalizedG2}) with $n=2$ to extract the antibunching and bunching parameters, as shown in Figs.~\ref{fig:simulations}(c)-(e).
For reasons explained later in this section, the simulated data were best described by an $n=2$ model despite having three eigenvalues.
See Materials and Methods for more information regarding the simulations. 

The four-level model consists of a ground state (level 1), an excited radiative state (level 2), a higher-lying excited state (level 3) and a nonradiative metastable state (level 4).
We consider two optical excitation pathways from the ground state to excited states 2 or 3, represented by the rates $\Gamma_{12}$ and $\Gamma_{13}$, respectively.
The magnitudes of these two rates depend on the corresponding optical cross-sections for absorption at the excitation wavelength.
A difference in cross section can result from the difference in electric dipole matrix elements between the different electronic states, the atomic configuration coordinate overlap for vibronic transitions, or both of these factors.
For the simulations in Fig.~\ref{fig:simulations}, we set $\Gamma_{12}=0$, since we are particularly interested in the situation where $\Gamma_{12}/\Gamma_{13}\ll 1$, such that the dynamics feature indirect excitation of the radiative state 2 \textit{via} nonradiative relaxation from excited state 3, at a rate $\kappa_{32}$.
This was informed by the nonlinear power-scaling of $\gamma_{1}$ for QEs A, B and C.
In the Supplementary Materials, we report simulations over a range of settings where $\Gamma_{12}/\Gamma_{13} \in [0,2]$, with qualitatively similar results (see Supplementary Figure S8).
 
In addition to the indirect excitation pathway formed by states 1, 2, and 3, optical excitation results in population and relaxation of metastable state 4 \textit{via} nonradiative transitions with rates $\kappa_{24}$ and $\kappa_{41}$.
We consider two types of nonradiative transition mechanism for the metastable state: spontaneous and optically pumped.
Spontaneous transition rates are independent of the optical excitation rate (in this case, $\Gamma_{13}$), whereas optically pumped transition rates scale linearly with $\Gamma_{13}$.
In this model, the optically pumped transition rates $\kappa_{24}$ and $\kappa_{41}$ can approximate more complicated processes; for example, they could involve re-pumping from levels $2\rightarrow3$ or from levels $4\rightarrow 3$ with subsequent nonradiative relaxation (see Supplementary Figure S10), or they could involve transient population of additional levels. 
Their approximation as individual pumped transitions remains accurate as long as optical pumping remains the rate-limiting step.
The key observable difference between spontaneous and optically pumped transitions manifests in the excitation power dependence of the corresponding bunching rate (Fig.~\ref{fig:simulations}d); the bunching rate for spontaneous transitions features a non-zero zero-power offset and saturates at high power, whereas the bunching rate for optically pumped transitions has zero offset and scales nearly linearly with power, even as the corresponding bunching amplitude (Fig.~\ref{fig:simulations}e) saturates.

For both bunching mechanisms, the simulated data were best described by only a single bunching level ($n=2$ in Eq.~\ref{eq:generalizedG2}) despite the fact that there should be 3 eigenvalues that describe this system.
The reason for this is that the indirect excitation and emission process through levels 1, 2, and 3 can lead to two of the eigenvalues being complex.
These eigenvalues have the largest real values and are responsible for the antibunching dynamics.
When we include practical limitations on timing resolution and signal-to-noise ratio at short delay times, the fit cannot distinguish these two values, and the goodness-of-fit analysis prefers a single real rate that approximates the true model.
The result is an effective antibunching rate that scales nonlinearly with increasing excitation rate.
This effect persists even when a direct transition from state $1\rightarrow 2$ is included.
We performed simulations varying the ratio $\Gamma_{12}/\Gamma_{13}$, and observed qualitatively similar results (see Supplementary Figure S8).

\section{Discussion}

\subsection{Photoluminescence, spectral, and polarization properties}

Our experiments provide clear evidence that visible QEs in h-BN occur as isolated point defects with emission originating from a single, dominant optical transition.
The QEs are spatially resolved in high signal-to-background $\micro$-PL images.
They exhibit PL saturation, high polarization visibility in emission, and emission spectra consistent with individual vibronic transitions.
Most convincingly, several emitters exhibit pure single-photon emission, with $\autocorr{0}=0$ within small experimental uncertainty, as shown in Fig.~\ref{fig:irfg2}.
This finding contrasts with previous suggestions that h-BN's QEs occur in pairs \cite{Bommer2019a}.
We do not contend, however, that such pairing cannot occur.
On the contrary, in the course of our experiments we observed multiple instances of spatially isolated emitters with high polarization visibility, and yet $\autocorr{0}$ is substantially larger than zero.
We have focused here on emitters showing the highest likelihood of being single defects.


Qualitatively, the QEs' room-temperature PL spectra are similar to those reported elsewhere in the literature \cite{Exarhos2017,Stern2019,Jungwirth2017,Jungwirth2016}.
The analysis shown in Fig.~\ref{fig:lineshapes} indicates that the PL spectra are consistent with individual vibronic transitions between two optical-dipole-coupled electronic states.
The one-phonon lineshapes for QEs A, B, C, and E are all qualitatively similar despite the fact that QEs A and B feature ZPL energies near 1.9~eV, compared to 2.1~eV for QEs C and E.
All four emitters exhibit strong coupling to low-energy phonons ($\lesssim$50~meV) as well as to higher-energy phonons (150-200~meV) that are typically associated with longitudinal optical modes in bulk h-BN \cite{Martinez2016a}.
Coupling to low-energy phonons is a key feature in determining the asymmetric shape of the dominant emission peak \cite{Wigger2019,Jungwirth2016}.
The ZPL corresponds to the transition from the lowest vibrational level of the initial (excited) state to the lowest vibrational level of the final (ground) state.  
When low-energy phonons are involved, transitions can occur from the lowest vibrational level of the initial state to the first vibrational level of the final state, showing up on the low-energy side of the ZPL and leading to the asymmetric shape.
Failure to account for low-energy phonons in interpreting experimental spectra leads to an underestimation of the Huang-Rhys factor, $S_\mathrm{HR}$, which quantifies the strength of the vibronic coupling and is a crucial parameter for comparing with theoretical calculations.
Our model captures the asymmetric spectral shape. 
However, the precise details of the low-energy phonon-coupling lineshape become correlated with the ZPL width (assumed to be Lorentzian) and $S_\mathrm{HR}$ when fitting the model to experimental data. 
We account for these correlations by performing the fits using varied constraints on the low-energy phonon coupling motivated by scaling considerations.
We follow the method described in Ref.~\cite{Exarhos2017}, in order to estimate uncertainties on $S_\mathrm{HR}$ and the ZPL linewidth.
Overall, we find that the best-fit ZPL linewidths are narrower or comparable to those reported in the literature for off-resonant excitation of h-BN's QEs at room temperature \cite{Dietrich2018,Dietrich2020,Akbari2021}, and the values of $S_\mathrm{HR}$ are somewhat higher.
We consider comparisons to theoretical proposals in detail later; we note here that the ZPL energies and $S_\mathrm{HR}$ values closely match the calculated properties of boron dangling bonds \cite{Turiansky2019a,Turiansky2021}.

The QE's polarization-resolved PL excitation and emission characteristics (Fig.~\ref{fig:summary}) begin to reveal more complicated features of their optical dynamics.
Both QEs A and C exhibit linearly polarized emission with nearly complete visibility, again consistent with emission through a single optical dipole transition.
For QE A, the PL intensity varies as a function of excitation polarization angle in a manner consistent with excitation through a single optical dipole, with high visibility and an angle aligned with the emission dipole, independent of excitation wavelength (532~nm or 592~nm).
In the case of QE C, the emission polarization visibility and dipole angle is similarly independent of the excitation wavelength.
However, QE C's excitation polarization dependence varies dramatically as a function of excitation wavelength; the excitation dipole is aligned with the emission under 592~nm excitation, but misaligned under 532~nm excitation with substantially reduced visibility.
Emission polarization data are not available for QEs B, D, and E, but the excitation polarization measurements are qualitatively similar to those for QEs A and C.
All three QEs show polarized absorption with varying degrees of visibility.

The heterogeneous polarization responses are consistent with previous observations for QEs in h-BN \cite{Jungwirth2016,Exarhos2017,Jungwirth2017,Ziegler2018}.
In particular, Ref.~\cite{Jungwirth2017} studied the variation of polarization visibility and alignment between excitation and emission as a function of the energy difference between the excitation photon energy and the ZPL photon energy, $\Delta E$.
They observed that the excitation and emission dipoles are aligned ($\Delta\theta=0$) when $\Delta E \lesssim \SI{200}{\milli\eV}$, whereas if $\Delta E \gtrsim \SI{200}{\milli\eV}$, $\Delta \theta$ can take any value.
Our observations are consistent with this empirical finding.
For QE A, the excitation and emission angles are aligned despite relatively large energy differences, $\Delta E^{592} = \SI{169}{\milli\eV}$ and $\Delta E^{532} = \SI{405}{\milli\eV}$, for 592~nm and 532~nm excitation, respectively.
For QE C, the dipoles are aligned for excitation at 592~nm ($\Delta E^{592} = \SI{10}{\milli\eV}$; $\Delta \theta^{592} = {0.0\pm1.4}{\degree}$) but misaligned at 532~nm ($\Delta E^{532} = \SI{247}{\milli\eV}$; $\Delta \theta^{532} = {46.5\pm1.6}{\degree}$).

Misalignment between absorption and emission dipoles is expected if the optical dynamics involve multiple excited states.
Whereas the invariance of the PL polarization, visibility, and spectral shape to exctiation energy implies that PL emission occurs through a single optical transition, off-resonant optical pumping can involve transient excitation of higher-lying excited states through transitions with different optical dipole orientations, which subsequently relax to the radiative state as shown in Fig.~\ref{fig:simulations}(a). 
Depending on energy level arrangement and the vibronic copuling strengths, a single excitation laser can drive both transitions between states $1\rightarrow2$ and $1\rightarrow3$.
The excitation polarization dependence will then reflect a superposition of two optical dipole transitions, with an orientation and visibility determined by the underlying dipole transition orientations and their relative optical cross section.
To test this hypothesis, we performed a simultaneous fit of QE C's emission and excitation polarization data under 532~nm excitation assuming a single shared dipole for excitation and emission \textit{via} states $1\leftrightarrow2$ together with a second dipole for excitation \textit{via} $1\rightarrow3$ (see Section S4 and Fig.~S2 in the Supplementary Materials). 
We find that the data are consistent with such a model, in which the dipole projection for transition $1\rightarrow3$ is misaligned from that of transition $1\leftrightarrow2$ by $63\pm1{\degree}$, and the ratio of excitation cross sections is $\Gamma_{12}/\Gamma_{13}\sim0.5$.

In interpreting these results, we note that the observation of highly polarized emission implies the presence of at least one symmetry axis for the underlying electronic states.
For most defect models under consideration, symmetry allows for optical dipole transitions aligned perpendicular to the h-BN plane (along $z$) or within the plane either parallel or perpendicular to the defect's symmetry axis (along $x$ or $y$).
Hence, the observation of dipoles misaligned by $\sim$60{\degree} seems surprising.
However, since our polarization-resolved experiments are primarily sensitive to the projection of the dipole perpendicular to the microscope's optical axis, it is possible that sample misalignment or local distortions of the defect that tend to tilt the $z$ axis could explain the observations.
Alternatively, our model of two superposed excitation dipoles might not capture all salient features of the excitation process; more than two transitions might be involved, and yet the superposition of any number of dipole absorption patterns will ultimately yield a polarization dependence consistent with Eq.~(\ref{eqn:dipole}).

\subsection{Optical dynamics}

PECS experiments reveal key details regarding the nature of the QEs' excited states and optical dynamics.
The PECS results summarized in Fig.~\ref{fig:photonStats} resolve individual dynamical processes, their associated timescales, and their dependence on optical excitation power. 
All QEs feature three or more timescales in their autocorrelation spectra, which implies that the optical dynamics involve at least four electronic levels.
In addition to antibunching on nanosecond timescales, all QEs exhibit bunching with two or more resolvable timescales that are orders of magnitude longer (typically microseconds to milliseconds).
These bunching timescales are broadly consistent with past observations \cite{Exarhos2017,Chejanovsky2016,Martinez2016a,Sontheimer2017,Boll2020,Stern2021}, and they indicate the role of metastable dark states in the optical dynamics.
Here, we emphasize and discuss two key features of the PECS measurements in Fig.~\ref{fig:photonStats}: the nonlinear power dependence of the antibunching rate, $\gamma_1$, that is clearly observed for QEs A, B, and C; and the heterogeneous behavior of the bunching rates and amplitudes, which feature qualitatively diverse power-dependent variations.

For a QE featuring a direct optical transition between a ground state and a radiative excited state, the antibunching rate scales linearly as a function of optical excitation power, with a zero-power offset corresponding to the QE's spontaneous emission rate.
This is the case even for QEs that also feature metastable charge and spin states, such as the NV center in diamond \cite{Doherty2013}. 
In Supplementary Figure S5, we present measurements of the antibunching rate of single NV centers in nanodiamonds as a function of excitation power; the results show clear linear scaling and a zero-power offset for $\gamma_1$ consistent with the expected optical lifetime.
The PECS observations of h-BN's QEs in Fig.~\ref{fig:photonStats} defy this expectation.
The power-scaling of $\gamma_1$ for QEs A, B, and C is clearly sublinear, with a saturation behavior (Model II or Model III) characterized by a steep slope at low power tapering off to a shallow slope at high power.
Moreover, the $\gamma_1$ measurements for QE A are all less or equal to the measured spontaneous decay rate (dashed line in the upper-left panel of Fig.~\ref{fig:photonStats}), whereas $\gamma_1$ always exceeds the spontaneous rate for a direct optical transition.
The zero-power offset for $\gamma_1$ in QEs A and B is consistent with zero but poorly constrained due to the steep low-power slope; the offset is non-zero for QEs C, D, and E.
QEs D and E exhibit linear power-scaling of $\gamma_1$, however the range of available powers is smaller than for the other emitters, and we cannot rule out a saturation behavior at higher power.
Previous studies of QEs in h-BN have revealed hints of power-independent antibunching rates \cite{Sontheimer2017} and nonlinear power scaling \cite{Chejanovsky2016,Boll2020}, however these observations were never satisfactorily explained.

The antibunching rate's nonlinear power dependence can be understood in the context of an indirect excitation mechanism, as illustrated in Fig.~\ref{fig:simulations}(a), where optical excitation leads to the population of multiple states: levels 2 and 3, with competing rates $\Gamma_{12}$ and $\Gamma_{13}$.
Indirect population of the radiative state (level 2) through such a mechanism creates a rate-limiting step ($3\rightarrow2$) to the optical emission pathway ($2\rightarrow1$) that leads to nonlinear scaling of the observed antibunching rate, as shown in Fig.~\ref{fig:simulations}(c).
The rate-limiting nature of this process is intuitively obvious in the limit where $\Gamma_{12}/\Gamma_{13} \ll 1$.
However, we find that the nonlinear saturation behavior remains qualitatively consistent independent of the pumping-rate ratio across a wide range of simulation settings where $\Gamma_{12}/\Gamma_{13} \in [0,2]$ (see Supplementary Figure S8).
In the regime $\Gamma_{12}/\Gamma_{13}\ll1$, the population of level 2 is still mostly determined by the indirect excitation pathway through level 3, with rate $\kappa_{32}$, and the dominant antibunching rate saturates to a value close to $\kappa_{32}+\Gamma_{21}$. 
In the regime where $\Gamma_{12}/\Gamma_{13}>1$, two underlying rates in the dynamical system are associated with the antibunching dip.
As discussed previously, the eigenvalues associated with these rates can be real or complex depending on the relative magnitudes of transitions in the system.
However, the fast rate associated with the direct population of level 2 and the subtle signatures of complex eigenvalues on the shape of the antibunching dip turn out not to be detectable when we include realistic assumptions for the experimental limits on timing resolution and shot noise.
Instead, we observe a single effective antibunching rate $\gamma_{1}$ that exhibits nonlinear saturation similar to the slow rate.

The bunching dynamics observed in Fig.~\ref{fig:photonStats} can also be understood within our optical dynamics models by including metastable shelving states.
In Fig.~\ref{fig:simulations}, the key qualitative difference between spontaneous population of the metastable state(s) (\textit{e.g.}, spin-dependent intersystem crossings) and optically pumped transitions (\textit{e.g.}, ionization/recombination) appears in the power-scaling and zero-power offset of the associated bunching rate.
Spontaneous transitions are characterized by a rate with a non-zero offset that tends to saturate with increasing pumping power, whereas optically pumped transitions have zero offset and increase quasi-linearly.
Previous studies considering the power-scaling of bunching rates for QEs in h-BN nano-flakes and exfoliated flakes have proposed similar optically-pumped models \cite{Boll2020,Chejanovsky2016}.
Similar behavior has also been observed in color centers such as the silicon-vacancy center in diamond, attributed to power-dependent de-shelving from higher lying states to the metastable state \cite{Neu2012PhotophysicsEmission}.
However, the heterogeneity and complexity of these processes for QEs in h-BN, both regarding the number of levels and the type of transitions, have not been considered before.

We observe both qualitative bunching behaviors in the data of Fig.~\ref{fig:photonStats}, with several QEs exhibiting multiple bunching levels that apparently have different transition mechanisms.
In some cases, individual bunching rates exhibit power scalings with features of both phenomena; for example, $\gamma_2$ for QEs A, B, D, and E appears to have a non-zero offset and yet increase linearly with power.  
This could indicate that the associated state can be populated both spontaneously and through an optically pumped pathway.
Our simulations support this intuitive reasoning.
For example, Supplementary Figure S9 shows the results of simulations of the same four-level system as in Fig.~\ref{fig:simulations}(a), but with rates chosen to reproduce the observations for QE A from Fig.~\ref{fig:photonStats}.
We indeed find that a combination of spontaneous and optically pumped transitions to the metastable state ($\kappa_{24}$ and $\kappa_{41}$) yields a bunching rate $\gamma_2$ with a non-zero offset that scales linearly with pumping power.
Moreover, setting $\kappa_{32}<\Gamma_{21}$ creates a situation where the spontaneous emission rate exceeds the observed antibunching rate, $\Gamma_{21}>\gamma_1$, over a wide range of pumping power, in agreement with our observations.
The quantitative magnitudes of $\gamma_1$, $\gamma_2$, and the bunching amplitude, $C_2$, are also reproduced by the simulations.
This highlights the versatility of optical dynamics simulations as a valuable tool to recreate or predict optical dynamics based on complex combinations of radiative and nonradiative processes.
To fully capture the observed dynamics of any particular QE, including the additional bunching rates $\gamma_3$ and $\gamma_4$ (where applicable), more metastable states are required in the simulations.
We further note that the number of observed bunching timescales represents a lower limit on the number of metastable states, and hence some states could actually represent multiplets associated with different spin manifolds. 
Even with those caveats, these observations present the opportunity for quantitative comparisons with theoretical predictions.

\subsection{Consistency with theoretical proposals}
Several defect structures have been proposed as the origin of visible-wavelength single-photon emission in h-BN, including the boron dangling bond (DB)~\cite{Turiansky2019a}, $V_{\rm N}N_{\rm B}$~\cite{Tran2016a}, $V_{\rm N}C_{\rm B}$~\cite{Sajid2020a}, and $V_{\rm B}C_{\rm N}$~\cite{Mendelson2020}.
The negatively-charged boron vacancy, $V_{\rm B}^-$, has been suggested to give rise to an optically detected magnetic resonance signal observed for emitter ensembles~\cite{Gottscholl2020}, however $V_{\rm B}^-$ has a ZPL of $\sim$1.7~eV and couples more strongly to phonons ($S_\mathrm{HR} \sim 3.5$)~\cite{Ivady2020}, producing a PL band between 800-900~nm that does not overlap with the emitters considered here.
Early studies highlighted $V_{\rm N}N_{\rm B}$ as the potential origin of visible QEs~\cite{Tran2016a}, but recent calculations show that the coupling to phonons is substantially larger than observations~\cite{Sajid2020a}.
More recently, $V_{\rm B}C_{\rm N}$ has been proposed based on the observation that carbon is correlated with the emission signal, but the calculated PL spectrum~\cite{Mendelson2020} does not match our observations.
The $V_{\rm B}C_{\rm N}$ calculations also predict a single, linearly-polarized absorption dipole, which is inconsistent with our measurements.
The calculated PL spectrum and strain dependence of $V_{\rm N}C_{\rm B}$~\cite{Sajid2020a} are in reasonable agreement with the our observations.
However, the optical transition for $V_{\rm N}C_{\rm B}$ occurs in the triplet channel, while the calculated ground state is a singlet;
the authors did not propose a mechanism through which the triplet channel is populated quickly enough to give rise to the optical emission they considered.

The boron DB is predicted to possess an optical transition at 2.06~eV with a Huang-Rhys factor of 2.3~\cite{Turiansky2019a}, which is in close agreement with the values observed in this study.
In addition, the variations in ZPL and $S_\mathrm{HR}$ for the observed emitters can be explained by out-of-plane distortions~\cite{Turiansky2021}.
The ground state of the boron DB is a singlet, and the predicted existence of a triplet excited state can explain the presence of level 4 in Fig.~\ref{fig:simulations}(a).
Another important feature of the boron DB model is the proximity of the states to h-BN's conduction band~\cite{Turiansky2019a};
this allows electrons to be optically excited directly into the conduction band, depending on the excitation energy, explaining the misalignment of the absorptive and emissive dipole when the excitation energy is increased.
Other proposed models do not provide an explanation for the misalignment.
For instance, in the case of $V_{\rm N}C_{\rm B}$ the optical transition occurs in the neutral charge state, and for the excitation energies considered here, photoionization will not occur~\cite{Wu2017}.

Within the boron DB model, we would interpret level 3 in Fig.~\ref{fig:simulations}(a) as the conduction band and $\kappa_{32}$ as the nonradiative capture rate.
To support this interpretation, we have estimated the relevant capture rate $\kappa_{32}$ of a photoionized electron from the conduction-band minimum into the DB excited state [level 2 in Fig.~\ref{fig:simulations}(a)].
This capture rate is a product of a capture coefficient and the density of electrons in the conduction band.
A first-principles calculation yields a capture coefficient of $4 \times 10^{-7}$~cm$^3$~s$^{-1}$ (see Supplementary Materials, Section S9).
The density of electrons is estimated based on the thermal velocity of the photoionized electron ($\sim10^5$~m~s$^{-1}$) and a typical electron energy relaxation time of $\sim1$~ps~\cite{Yan2014}.
In the time it takes the electron to relax to the conduction-band minimum, it can thus travel $\sim100$~nm;
this distance corresponds to an effective electron density of $2.4 \times 10^{14}$~cm$^{-3}$.
Multiplying this value with the calculated capture coefficient gives a rate of $\kappa_{32}\sim100$~MHz, in compelling agreement with the observed saturation antibunching rates of $\gamma_1\sim 300$-800~MHz for QEs A, B, and C.
Our calculations also show that capture into the excited state is favored over capture into the ground state by more than 5 orders of magnitude, justifying the neglect of $\kappa_{31}$ in the general model of Fig.~\ref{fig:simulations}(a).

The inclusion of photoionization allows us to further rationalize the heterogeneity in bunching behavior of the observed emitters:
the photoionized electron is not necessarily re-captured at the same QE, but may instead be captured by a neighboring defect, leaving the QE in a nonfluorescent, ionized configuration that likely requires optical excitation of additional free electrons to restore emission \textit{via} subsequent electron capture.
This process would be represented in Fig.~\ref{fig:simulations}(a) by an optically pumped transition, where level 4 represents an ionized state of the QE.
The emitters may therefore be highly sensitive to the local defect environment.
Unlike other proposed defect models, we conclude that the boron DB model is thus capable of explaining numerous aspects of the experimental observations, lending support to this proposed microscopic structure.

\section{Conclusion}

The observations in this work reveal that h-BN's QEs have intricate electronic level structures and complex optical dynamics including multiple charge or spin manifolds.
Our proposed electronic-structure models complement previous reports \cite{Sontheimer2017,Chejanovsky2016,Boll2020} and explain the quantitative features of our observations.
In particular, the models explain the observation of nonlinear power-scaling of the antibunching rate as well as heterogeneous magnitudes and power-scaling behavior of multiple bunching rates.
Whereas past reports have lacked consensus on mechanisms to explain the observed optical dynamics of h-BN's QEs, and many posited chemical and electronic structure models have failed to adequately explain the heterogeneous observations, we show that the boron dangling bond model is remarkably consistent with experiments, especially accounting for the role of local distortions, photoionization, electron capture, and the QEs' heterogeneous local defect environment.
Future experiments should be designed to investigate these details, for example time-domain studies of transients associated with charge and spin dynamics, and temperature- and excitation-energy-dependent variations of the PL lineshape, vibronic spectrum, and polarization-dependent excitation cross section.
Combined with theoretical models, such experiments can resolve the underlying transition rates and resolve the disparate influences of the QEs' intrinsic properties with those of their local environments.
The observation of pure single-photon emission with $\autocorr{0}=0$ resolves earlier questions about h-BN's QEs \cite{Bommer2019a}, affirming their potential for use in photonic quantum technologies.
More generally, we hope that the approach and techniques presented in this work\,---\,especially the quantitative use of PECS\,---\,present a model to formulate optical dynamics models for QEs in any material platform \cite{Fishman2021,Bassett2019}.
Our models can be adapted to account for recent observations of magnetic-field-dependent optical dynamics \cite{Exarhos2019} and optically detected magnetic resonance \cite{Chejanovsky2021,Stern2021} in h-BN.
Subsequently, they can be used to design protocols for initialization, control, and readout of quantum-coherent spin states for quantum information processing and quantum sensing.

\section{Materials and Methods}
\subsection{Sample preparation}

We used a confocal microscope (see Supplementary Materials) to isolate individual QEs in h-BN under ambient conditions. 
The h-BN samples were sourced from two batches (purchased $\sim$2 years apart) of bulk, single crystals from HQ Graphene.
Each batch consisted of roughly 20 different individual crystals.
The bulk crystals were mechanically exfoliated using a dry transfer process \cite{Huang2015} resulting in thin ($\SI{\leq100}{\nano\meter}$) and large area ($\SI{\sim10}{\micro\meter}$) flakes of h-BN.
The exfoliated flakes were transferred to a SiO$_{2}$/Si substrate with micro-fabricated circular trenches \SIrange{4}{8}{\micro\m} in diameter and \SI{5}{\micro\m} deep  \cite{Exarhos2017} using a dry transfer process.
Table S1 in Supplementary Materials highlights the crystal from which the h-BN flake under study came.

Prior to the optical studies, the exfoliated h-BN samples were cleaned with a soft O$_2$ plasma (Anatech SCE 106 Barrel Asher, \SI{50}{\W} of power, \SI{50}{\sccm} O$_2$ flow rate) for 5 minutes to remove polymer residues resulting from the transfer process. 
The samples were then annealed in a tube furnace at \SI{850}{\celsius} in low flow Ar atmosphere for 2 hours. 
Annealing h-BN has been found to brighten the emitters \cite{Breitweiser2020}.
While not the focus of study here, annealing for longer time (2 hours vs commonly used 30 minutes) appears to improve emitter stability.
One sample was also exposed to a focused ion beam chamber operated in scanning electron mode (FEI Strata DB235 FIB SEM) but was not directly exposed to the electron beam.
Supplementary Table S1 summarizes the three samples investigated, the QEs studied in each sample and the annealing treatment received by each sample.

\subsection{Experimental details}

Supplementary Figure S1 depicts a simplified schematic of the room-temperature confocal microscope used to measure the QEs.
There are two available excitation sources: a \SI{532}{\nano\meter} (green) cw laser (Coherent, Compass 315M-150) and a \SI{592}{\nano\meter} (orange) cw laser (MPB Communications, VF-P-200-592).
The power and polarization of each excitation path can be independently selected.
Reported power values are measured just prior to the objective.
In addition, a shutter completely blanks the excitation source when imaging is not in use to mitigate unnecessary light exposure.
The excitation paths are combined with the collection path using a long pass (LP) dichroic mirror (Semrock, BrightLine FF560-FDi01 for green and Semrock, BrightLine FF640-FDi01 for orange).
The LP dichroic cut-off is 560 nm for green excitation and 640 nm for orange excitation. 
A fixed half-wave plate in each of the excitation paths corrects for the birefringence induced by the dichroic mirrors.
The co-aligned excitation and collection paths are sent through a $4f$ lens system with a fast steering mirror (Optics in Motion, OIM101) and a 0.9 NA 100x objective (Olympus, MPI Plan Fluor) at the image planes.
This allows for the collection of wide-field, rastered, micro-photoluminescence ($\micro$-PL) images.
The objective is mounted on a stage system for changing the field of view.

The collection path consists of a linear polarizer (Thorlabs, WP25M-VIS) for measuring the emission polarization as well as a wide-band variable retarder (Meadowlark, LRC-100) which compensates for the birefringence induced by the dichroic.
A LP filter specific to the excitation color fully extinguishes any scattered excitation light and the Raman signal.
The cut-on wavelengths are \SI{578}{\nano\meter} (Semrock, BLP01-568R-25) and \SI{650}{\nano\meter} (Semrock, BLP01-635R-25) for green and orange, respectively.
The filtered light is focused onto the core of a \SI{50}{\micro\meter} core multi-mode fiber (Thorlabs, M42L01) acting as a pinhole.
The output of the fiber is connected to a fiber switch (DiCon, MEMS 1x2 Switch Module) which can switch the collected emission to either a 50:50 visible fiber splitter (Thorlabs, FCMM50-50A-FC) or a spectrometer (Princeton Instruments, IsoPlane160 and Pixis 100 CCD).
The outputs of the fiber splitter are sent to two identical single-photon counting modules (SPCM, Laser Components, Count T-100) resulting in a Hanbury Brown Twiss interferometer.
The outputs of the SPCMs are either measured by a data acquisition card (National Instruments, DAQ6323) for general-purpose counting or a time-correlated single-photon counting (TCSPC) module (PicoQuant, PicoHarp 300) for recording the photon time-of-arrival information with a full system resolution of \SI{\sim350}{\pico\second}.

\subsection{Spectra analysis}
The PL spectra are collected as multiple exposures and averaged after correcting for dark counts, cosmic rays and wavelength-dependent photon collection efficiency.
The PL spectra are measured as a function of wavelength, $\lambda$ and binned to determine spectral distribution function, $S(\lambda)$.
To analyze the vibronic coupling, the measured spectra must be converted to a form suitable for analysis with the general theory of electron-phonon coupling in three dimensional crystals \cite{Maradudin1966,Davies1974}.
To do this, the spectral probability distribution function is obtained through
\begin{equation}
    S(E) = S(\lambda)\frac{hc}{E^{2}}
    \label{eq:S(E)}
\end{equation}
where $h$ is Planck's constant, $c$ is the speed of light, and $E$ is the photon energy.
The emission lineshape, $L(E)$ is derived from $S(E)$ as 
\begin{equation}
    L(E) = \frac{S(E)}{E^3}
    \label{eq:L(E)}
\end{equation}
which accounts for the photon-energy dependence of spontaneous emission.
The emission lineshape is then fit  following the method described in \cite{Exarhos2017}. 
From the fit, the following free parameters are determined: the ZPL energy, $E_\mathrm{ZPL}$, the ZPL Lortentzian linewidth, $\Gamma_\mathrm{ZPL}$, the Huang-Rhys factor, $S_\mathrm{HR}$, and the one-phonon vibronic coupling lineshape, approximated as an interpolated vector of values spanning the phonon spectrum in h-BN. 
The Debye-Waller factor, $w_\mathrm{DW}$, can be calculated from $w_\mathrm{DW}=e^{-S_\mathrm{HR}}$.

\subsection{Second-order photon autocorrelation function}

For a given QE, all autocorrelation measurements are performed with the excitation polarization set at the angle of maximum excitation and the collection path has the polarizer removed.
Due to the varying QE brightness, which affects the signal-to-noise ratio of the antibunching signal, measurements are integrated for 10 s to 140 min with repositioning occurring every 2 min.
All errors from the fitting denote one standard deviation.

Due to timing jitter in the single photon counting modules (SPCMs) introducing systematic artifacts at short delay times, the data are analyzed in two stages: logarithmic and linear scales.
First, the the autocorrelation data are binned over a log scale for visualizing the dynamics over 9 orders of magnitude in time (0.1 $ns$ to 1 $s$) corrected for background \cite{Brouri2000} (see Supplementary Materials), and then fit by multiple instances of Eq.~\ref{eq:generalizedG2} with $n=[2, 5]$.
The best fit, and corresponding $n$, is then determined by calculating the AIC and comparing the reduced chi-squared statistic.
This method determines the number of bunching levels and their rates and amplitudes that best explain the observations.
The QE's autocorrelation data are then binned over a linear scale that contains the antibunching features ($\tau \le \SI{30}{\nano\second}$). 
To account for the timing jitter in the SPCMs, the instrument response function (IRF) is found by measuring the autocorrelation signal of an attenuated picosecond pulsed laser sent through the HBT interferometer and binned over the same linear scale as the QE.
A convolution of the IRF with a modified Eq.~\ref{eq:generalizedG2} is fit to the background-corrected data, given by
\begin{equation}
    \tilde{g}^{(2)}(\tau) = \mathrm{IRF} * (1 - C_1 e^{-\gamma_1|\tau|} + C_\mathrm{B}(\tau))
    \label{eq:linear_g2}
\end{equation}
where $C_\mathrm{B}(\tau)$ is the total bunching contribution found from the log scale analysis (first step) and only $C_1$ and $\gamma_1$ are allowed to vary.
The autocorrelation at zero delay is then given by
\begin{equation}
    \tilde{g}^{(2)}(0) = 1 - C_1 + \sum_{i=2}^n C_i
    \label{eq:g2_0}
\end{equation}
which is used to determine the purity of single-photon emission from the QE.

\subsection{Electronic level structure simulations}

A four-level optical rate equation is used to model aspects of the observed autocorrelation data. 
The model is defined as
\begin{align}
    \mathbf{\dot{P}} = \mathbf{G}\mathbf{P}
    \label{eq:masterEq}
\end{align} 
where $\mathbf{P}$ is a vector of state populations $P_i$ and $\mathbf{G}$ is a generator matrix describing the transition rates and is given by
\begin{align}
    \bf{G} = \
    \begin{pmatrix}
    -\Gamma_{13} & \Gamma_{21}  & 0 & \kappa_{41} \\
    0 & - \Gamma_{21} - \kappa_{24}  & \kappa_{32} & 0 \\
    \Gamma_{13} & 0 & -\kappa_{32} & 0 \\
    0 &  \kappa_{24}  & 0 & -\kappa_{41} 
    \end{pmatrix}
    \label{eq:generatorMatrix}
\end{align}
where $\Gamma_{13}$ is the excitation rate, $\Gamma_{21}$ is the radiative emission rate, and $\kappa_{ij}$ are nonradiative rates that are either fixed or a proportion of the excitation rate.
The $\autocorr{\tau}$ can be found by calculating the time evolution of the radiative state, $P_2$ given the system started in state $P_1$ and normalizing by the steady state population of $P_2$.
This is given by
\begin{align}
    \autocorr{\tau} = \frac{P_2(t_2 | P(t_1) = (1, 0, 0, 0))}{P_2(\infty)}
    \label{eq:simulatedG2}
\end{align}
where $\tau=t_2-t_1$.
The differential equation (Eq.~\ref{eq:masterEq}) given the initial state is solved in MATLAB using the function \texttt{ode15s}.
Timing resolution limitations and shot noise are added to the simulated autocorrelation function to best recreate the measurements.
To model timing resolution, the simulated data are only analyzed for $t_0 \ge \SI{0.5}{\nano\second}$.
To include shot noise, a standard deviation, $\sigma_0$, is set for the first delay time.
Assuming shot noise, this standard deviation is converted to mean number of photons as 
\begin{align}
    \braket{N_0} = \sigma_0^{-2}
\end{align}
The log-scale processing results in the average number of photon correlations detected in each bin increasing linearly with the delay time,
\begin{align}
    \braket{N(\tau)} = \braket{N_0}\frac{\tau}{\tau_0}
\end{align}
From this, a simulated, noisy $\autocorr{\tau}$ is calculated as
\begin{align}
    \autocorr{\tau}_\mathrm{Noisy} = \frac{\mathrm{Poiss}(\autocorr{\tau}\braket{N(\tau)})}{\braket{N(\tau)}}
\end{align}
where Poiss is a Poission distribution.
The simulated autocorrelation data are analyzed with the same fitting framework as the measured data.
The general model parameters are as follows:
        $\Gamma_{21}=\SI{300}{\mega\hertz}$,
        $\Gamma_{13} = a\Gamma_{21}$ where $a=[.01, 10]$,
        $\Gamma_{12} = x\Gamma_{13}$ where $x=[0, 2]$,
        $\kappa_{32}=\SI{600}{\mega\hertz}$.
        For the spontaneous bunching,
        $\kappa_{24}=\SI{60}{\kilo\hertz}$ and $\kappa_{41}=\SI{30}{\kilo\hertz}$.
        For the pumped bunching,
        $\kappa_{24}=\SI{6}{\kilo\hertz\per\mega\hertz}\times\Gamma_{13}$ and $\kappa_{41}=\SI{3}{\kilo\hertz\per\mega\hertz}\times\Gamma_{13}$.

\subsection{Theoretical calculations}
We perform first-principles density-functional theory calculations as implemented in the VASP code~\cite{Kresse1996,Kresse1996a}.
We utilize the hybrid functional of Heyd, Scuseria, and Ernzerhof~\cite{Heyd2003,HeydErratum2006} to ensure accurate energetics, electronic structure, and atomic geometries.
The fraction of Hartree-Fock exchange is set to 40\%, consistent with previous studies~\cite{Turiansky2019a,Turiansky2021}.
A plane-wave basis with projector augmented-wave potentials~\cite{Blochl1994} is used, and the energy cutoff for the basis is set to 520~eV.

The boron dangling bond is modeled in a 240-atom supercell with volume 2110~{\AA}$^3$ within periodic boundary conditions~\cite{Freysoldt2014}.
A single, special {\bf k}-point (0.25, 0.25, 0.25) is used to sample the Brillouin zone.
Lattice vectors are held fixed while the atomic forces are relaxed to below 0.01~{eV/\AA}.
To calculate the nonradiative capture coefficient, we utilize the formalism of Ref.~\cite{Alkauskas2014} implemented in the Nonrad code~\cite{Turiansky2021b}.

\section{Acknowledgements}

This work was primarily supported by the National Science Foundation (NSF) award DMR-1922278. 
J.A.G. was supported by an NSF Graduate Research Fellowship (DGE-1845298).
The authors gratefully acknowledge use of facilities and instrumentation in the Singh Center for Nanotechnology at the University of Pennsylvania, supported by NSF through the National Nanotechnology Coordinated Infrastructure (NNCI; Grant ECCS-1542153) and the University of Pennsylvania Materials Research Science and Engineering Center (MRSEC; DMR-1720530).
M.E.T. was supported by the NSF through Enabling Quantum Leap: Convergent Accelerated Discovery Foundries for Quantum Materials Science, Engineering and Information (Q-AMASE-i; DMR-1906325).
Computational resources were provided by the Extreme Science and Engineering Discovery Environment (XSEDE), supported by the NSF (ACI-1548562).
We gratefully acknowledge fruitful discussions with A. Alkauskas.

\subsection*{Author Contributions}

R.N.P. and L.C.B. designed the study. 
R.N.P. carried out the experiments. 
R.N.P., D.A.H., and L.C.B. analyzed and interpreted the experimental data and performed optical dynamics simulations. 
J.A.G. and L.C.B. carried out the PL spectra analysis and vibronic coupling calculations. 
R.E.K.F. contributed to the analysis and interpretation of photon emission correlation spectroscopy experiments.
M.E.T. and C.G.V.W. performed theoretical calculations of defect models.
R.N.P. and B.P. prepared the samples. 
T.-Y.H. and D.A.H. contributed to designing the experiment setup and data acquisition software. 
R.N.P., D.A.H., M.E.T., and L.C.B. wrote the manuscript.  
All authors contributed to discussions and commented on the manuscript.



\bibliography{arxiv_main}

\end{document}


\title{Supplementary Materials:\\
Probing the Optical Dynamics of Quantum Emitters in Hexagonal Boron Nitride}

\author{Raj N. Patel}
\affiliation{
Quantum Engineering Laboratory, Department of Electrical and Systems Engineering, University of Pennsylvania, Philadelphia, PA 19104, United States
}

\author{David A. Hopper}
\thanks{
Present address: MITRE Corporation, 7515 Colshire Dr.
McLean, VA 22102, USA
}
\affiliation{ 
Quantum Engineering Laboratory, Department of Electrical and Systems Engineering, University of Pennsylvania, Philadelphia, PA 19104, United States
}

\author{Jordan A. Gusdorff}
\affiliation{
Quantum Engineering Laboratory, Department of Electrical and Systems Engineering, University of Pennsylvania, Philadelphia, PA 19104, United States
}
\affiliation{
Department of Materials Science and Engineering, University of Pennsylvania, Philadelphia, PA 19104, USA
}

\author{Mark E. Turiansky}
\affiliation{
Department of Physics, University of California, Santa Barbara, CA 93106, United States
}

\author{Tzu-Yung Huang}
\affiliation{ 
Quantum Engineering Laboratory, Department of Electrical and Systems Engineering, University of Pennsylvania, Philadelphia, PA 19104, United States
}

\author{Rebecca E. K. Fishman}
\affiliation{ 
Quantum Engineering Laboratory, Department of Electrical and Systems Engineering, University of Pennsylvania, Philadelphia, PA 19104, United States
}
\affiliation{
Department of Physics and Astronomy, University of Pennsylvania, Philadelphia, PA 19104, USA
}

\author{Benjamin Porat}
\thanks{
Present address: Raytheon Intelligence and Space, 2000 E El Segundo Blvd, El Segundo, CA 90245
}\affiliation{
Quantum Engineering Laboratory, Department of Electrical and Systems Engineering, University of Pennsylvania, Philadelphia, PA 19104, United States
}

\author{Chris G. Van de Walle}
\affiliation{
Materials Department, University of California, Santa Barbara, CA 93106, United States
}

\author{Lee C. Bassett}
\email[Corresponding author.  Email: ]{lbassett@seas.upenn.edu}
\affiliation{ 
Quantum Engineering Laboratory, Department of Electrical and Systems Engineering, University of Pennsylvania, Philadelphia, PA 19104, United States
}

\date{\today} 


\maketitle

\clearpage

\subsection*{S1. Experiment setup}

\begin{figure*}[htb!]
    \renewcommand\figurename{Figure S}
    \let\nobreakspace\relax
    \setcounter{figure}{0}
    \centering
    \includegraphics[width=7in]{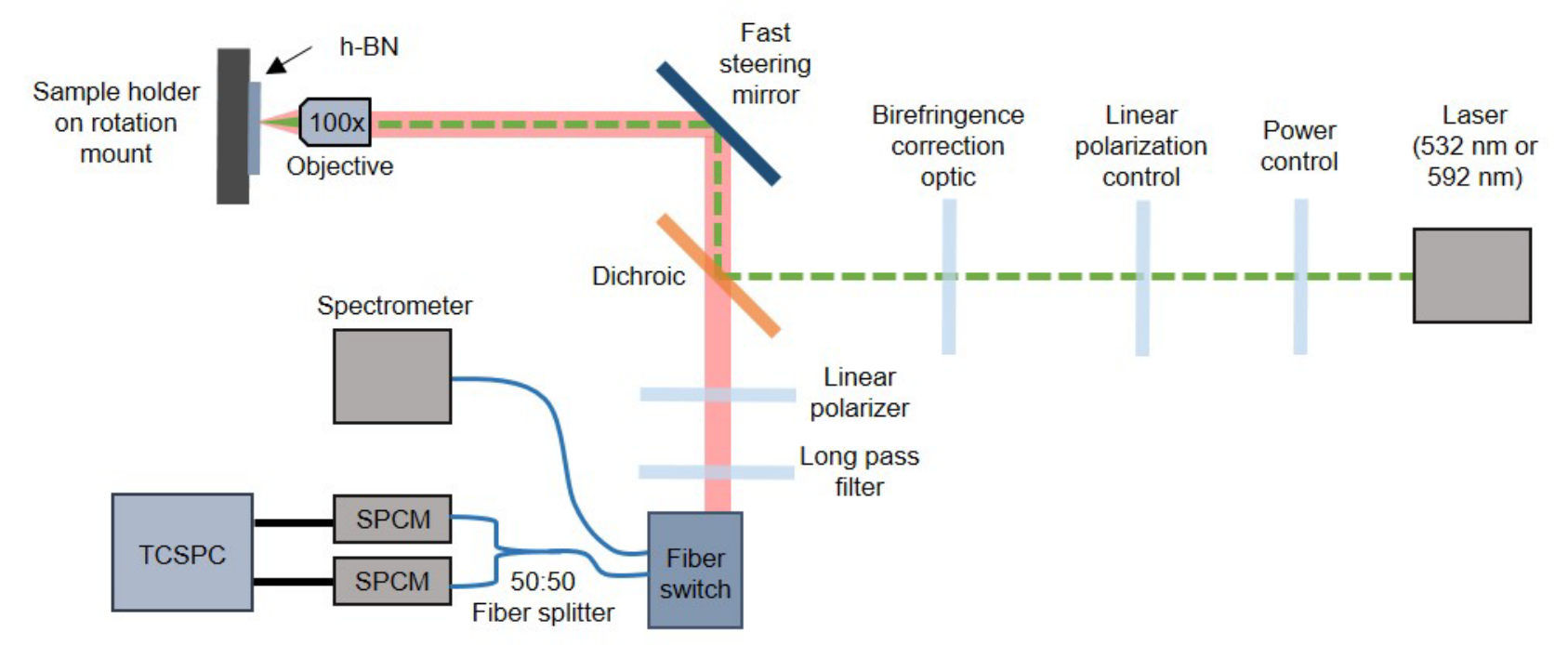}
    \caption{\textbf{Experiment Setup.} A simplified version of the room temperature optical setup showing the essential optical and electronic components used to probe the QEs in h-BN.
    The green dashed line represents the 532 nm (green) excitation path which can be switched to 592 nm (orange) excitation.}
    \label{fig:setup}
\end{figure*}

\subsection*{S2. Sample Information}

\begin{table*}[h!]
    \renewcommand{\thetable}{S\arabic{table}}
    \renewcommand\tablename{Table}
    \caption{\textbf{The Samples, Quantum Emitters and Sample Treatments.}}    
    \label{tab:samplePrep}
    \def\arraystretch{1.75}
    \begin{tabularx}{1\textwidth}
    { | X | X | X | X | }
    \rowcolor{lightGray}[\tabcolsep]
    \hline
    \textbf{Sample}$^1$ & \textbf{\RN{1}} & \textbf{\RN{2}} & \textbf{\RN{3}} \\ 
    \hline
    \textbf{h-BN crystal used for exfoliation}$^{2}$ &
    1 & 1 & 2 \\
    \hline
    \textbf{Quantum emitters} & 
    A & B & C, D $\&$ E \\
    \hline
    \textbf{Pre-annealing treatment} &
    Plasma cleaned at 50 W in 50 sccm O$_2$ for 5 minutes & 
    Plasma cleaned at 50 W in 50 sccm O$_2$ for 5 minutes &
    Plasma cleaned at 50 W in 50 sccm O$_2$ for 5 minutes \\
    \hline
    \textbf{Annealing treatment} &
    850 $^o$C for 2 hours in low pressure Ar atmosphere & 
    1. 850 $^o$C for 1 hour in low pressure Ar atmosphere \par
    2. Sample was in scanning electron microscope chamber but not directly exposed to e-beam$^{3}$ \par
    3. 850 $^o$C for 2 hours in low pressure Ar atmosphere &
    850 $^o$C for 2 hours in low pressure Ar atmosphere \\
    \hline    
    \end{tabularx}
\end{table*}

$^{1}$ Sample represents different substrates.

$^{2}$ Crystal 1 and 2 represent crystals in different orders purchased from HQ Graphene, $\sim$2 years apart.

$^{3}$ The QE was not found post treatment 1 and 2. It was only found post treatment 3.

\clearpage
\subsection*{S3: PL Characterization Summary}

\begin{table*}[ht]
    \renewcommand{\thetable}{S\arabic{table}}
    \renewcommand\tablename{Table}
    \caption{\textbf{Steady-State PL as a Function of Power, PL Spectra Analysis and Optical Polarization properties.}}
    \label{tab:plSummary}
    \def\arraystretch{1.75}
    \begin{tabularx}{1\textwidth}
    { | X | X | X | X | X | X | }
    \rowcolor{lightGray}[\tabcolsep]
    \hline
    \textbf{Quantum} \par \textbf{Emitter} &
    \textbf{A} & \textbf{B} & \textbf{C}$^1$ & \textbf{D}$^2$ & \textbf{E} \\
    \hline
    $\mathbf{C_{sat}^{592}(kCts/s)}$ &
    $84.4\pm5.1$ &
    $170.1\pm8.4$ &
    $323.0\pm38.3$ &
    $57.2\pm4.1$ &
    N/A \\
    \hline
    $\mathbf{P_{sat}^{592}(\micro W)}$ &
    $100.1\pm18.7$ &
    $124.2\pm18.0$ &
    $424.6\pm124.3$ &
    $126.2\pm22.54$ &
    N/A \\
    \hline
    $\mathbf{E_{ZPL}(eV)}$ & 1.927 & 1.871 & 2.086 & N/A & 2.1 \\
    \hline
    $\mathbf{S_{HR}}$ & 2.79 & 2.35 & 2.48 & N/A & 2.12 \\
    \hline
    $\mathbf{w_{DW}}$ & 0.06 & 0.095 & 0.084 & N/A & 0.12 \\
    \hline
    $\mathbf{\Gamma_{ZPL}(meV)}$ & 4.9 & 4.7 & 4.2 & N/A & 9.2 \\
    \hline
    $\mathbf{\theta_{ex}^{532}(deg)}^3$ &
    $168.17\pm0.97$ & N/A & $64.54\pm1.34$ & N/A & $108.77\pm1.26$ \\
    \hline
    $\mathbf{V_{ex}^{532}(\%)}^3$ & $83.6\pm1.4$ & N/A & $38.7\pm1.2$ & N/A & $56.1\pm1.4$ \\
    \hline
    $\mathbf{\theta_{ex}^{592}(deg)}^3$ &
    $169.71\pm2.76$ & $77.66\pm0.81$ & $118.29\pm1.18$ & $53.23\pm2.02$ & N/A \\
    \hline
    $\mathbf{V_{ex}^{592}(\%)}^3$ & $100.0\pm5.7$ & $80.88\pm1.7$ & $83.0\pm2.9$ & $57.6\pm2.3$ & N/A \\
    \hline
    $\mathbf{\theta_{em}^{592}(deg)}^3$ &
    $173.83\pm1.47$ & N/A & $118.29\pm0.66$ & N/A & N/A \\
    \hline
    $\mathbf{V_{em}^{592}(\%)}^3$ & $88.9\pm2.4$ & N/A & $90.5\pm1.9$ & N/A & N/A \\
    \hline
    \end{tabularx}
\end{table*}

$^{1}$ $\mathrm{C_{sat}^{532}=1457.7\pm98.1}$ kCts/s,  $\mathrm{P_{sat}^{532}=752.7\pm81.9}$ $\mathrm{\micro W}$, $\mathrm{\theta_{em}^{532}=111.09\pm0.81^o}$, $\mathrm{V_{em}^{532}=82.9\pm1.9\%}$.

$^{2}$ PL spectral information is incomplete. ZPL cutoff by the filter.

$^{3}$ Excitation ($\theta_{ex}$) and emission ($\theta_{em}$) dipole orientation, excitation ($V_{ex}$) and emission ($V_{em}$) visibility.

\subsection*{S4. Polarization Properties}

\subsubsection*{\textbf{Simultaneous Fit of Emission and Excitation Polarization of QE C for 532 nm Excitation}}

The emission and excitation polarization of QE C is independently modeled using Eq. 2 and 3 to estimate its optical dipole orientation and visibility.
Further, the misalignment of the emission and excitation dipole is estimated from the independent fit results using Eq. 4.
However, the low visibility of excitation polarization for 532 nm excitation suggests superposition of multiple excitation dipoles which means the excitation pathways could be via multiple dipoles, one of them oriented along the emission dipole.
To check this hypothesis, the following empirical equations are simultaneously fit to the emission and excitation polarization data:
\begin{equation}
    \renewcommand{\theequation}{S\arabic{equation}}
    \let\nobreakspace\relax
    \setcounter{equation}{1}
    I_{1} = A_{1}\cos^2{(\theta-\theta_{1})} + B_{1}
\end{equation}
\begin{equation}
    \renewcommand{\theequation}{S\arabic{equation}}
    I_{2} = A_{2}\cos^2{(\theta-\theta_{1})}+(1-A_{2})\cos^2{(\theta-\theta_{2})}+B_{2}
\end{equation}
where $A$ is the normalized amplitude, $B$ is the normalized offset, $\theta_{1}$ is the emission dipole orientation and orientation of one of the two excitation dipoles and $\theta_{2}$ is the second excitation dipole. 

\begin{figure*}[htb!]
  \renewcommand\figurename{Figure S}
  \let\nobreakspace\relax
  \centering
  \includegraphics[width=3.375in]{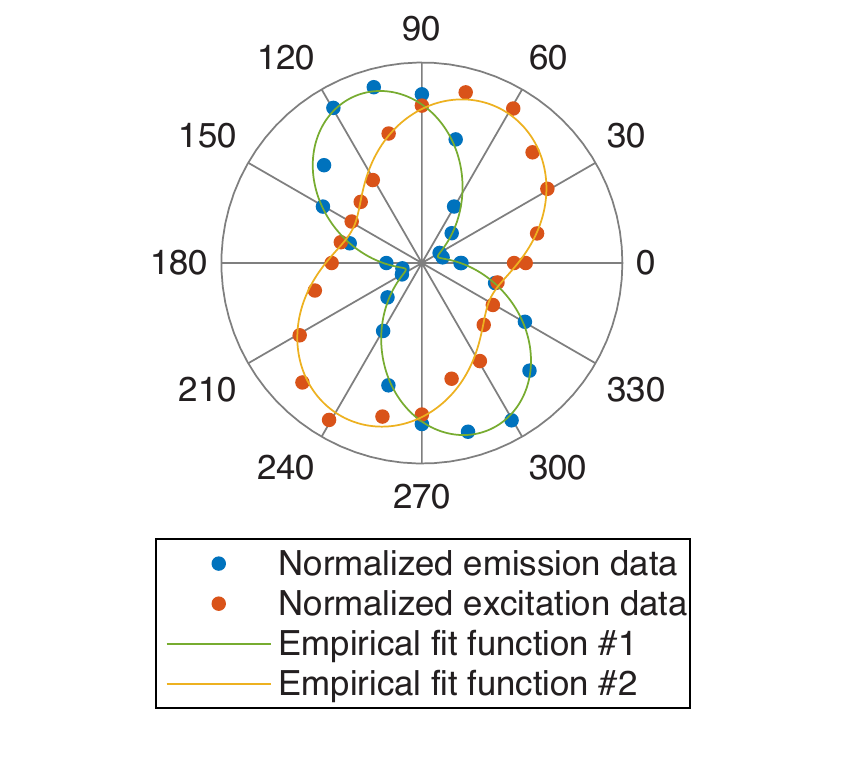}
  \caption{\textbf{Quantum emitter C: emission and excitation polarization with green excitation.}
  The normalized emission polarization is shown with blue circles and normalized excitation polarization is shown with orange circles.
  The green and yellow curves are simultaneous fits to the normalized emission and excitation polarization, respectively.
  }
  \label{fig:simultaneousDipoleFit}
\end{figure*}

Figure S\ref{fig:simultaneousDipoleFit} shows the normalized emission and excitation polarization data and the resultant simultaneous fits.
Table~\ref{tab:simultaneousFitResult} presents the fit result.
The emission and first excitation dipole is oriented $111.22^{o}$. 
This agrees with the independent fit to the emission polarization using Eq. 2 which gives dipole orientation of $111.09^{o}$ (Table~\ref{tab:plSummary}).
This proves the hypothesis that the excitation polarization is a superposition of two excitation dipoles, one aligned with the emission dipole and the other misaligned by $\sim$60$^{o}$.
Collectively, the two excitation dipoles result in an effective dipole with low visibility.

\begin{table*}[h!] 
    \renewcommand{\thetable}{S\arabic{table}}
    \renewcommand\tablename{Table}
    \caption{\textbf{Result of Simultaneous Fit of QE C's Emission and Excitation Polarization.}}
    \label{tab:simultaneousFitResult}
    \def\arraystretch{1.75}
    \begin{tabularx}{0.675\textwidth}
    { | p{4.75cm} | X | }
    \hline
    Emission ($I_{1}$) & 0.89 $\cos^2{(\theta-111.2)}$ + 0.09 \\
    \hline
    Excitation ($I_{2}$) & 0.34 $\cos^2{(\theta-111.2)}$ + 0.66 $\cos^2{(\theta-47.8)}$ + 0.19 \\ 
    \hline
    Dipole 1 ($\theta_{1}$) & $111.22\pm0.63^{o}$  \\
    \hline
    Dipole 2 ($\theta_{2}$) & $47.77\pm0.95^{o}$ \\
    \hline
    Emission amplitude ($A_{1}$) &  $0.89\pm0.02$ \\
    \hline
    Emission background ($B_{1}$) & $0.09\pm0.01$ \\
    \hline
    Proportion along 111.22$^{o}$ $(A_{2})$ & $0.34\pm0.01$ \\
    \hline
    Proportion along 47.77$^{o}$ $(1-A_{2})$ & $0.66\pm0.01$ \\ 
    \hline
    Excitation background ($B_{2}$) & $0.19\pm0.01$ \\
    \hline
    \end{tabularx}
\end{table*}

\subsection*{S5. Autocorrelation}

\subsubsection*{S5.1 Logarithmic Time Scale Autocorrelation Function}

\begin{figure*}[hb!]
  \renewcommand\figurename{Figure S}
  \let\nobreakspace\relax
  \centering    
  \includegraphics[width=6.75in]{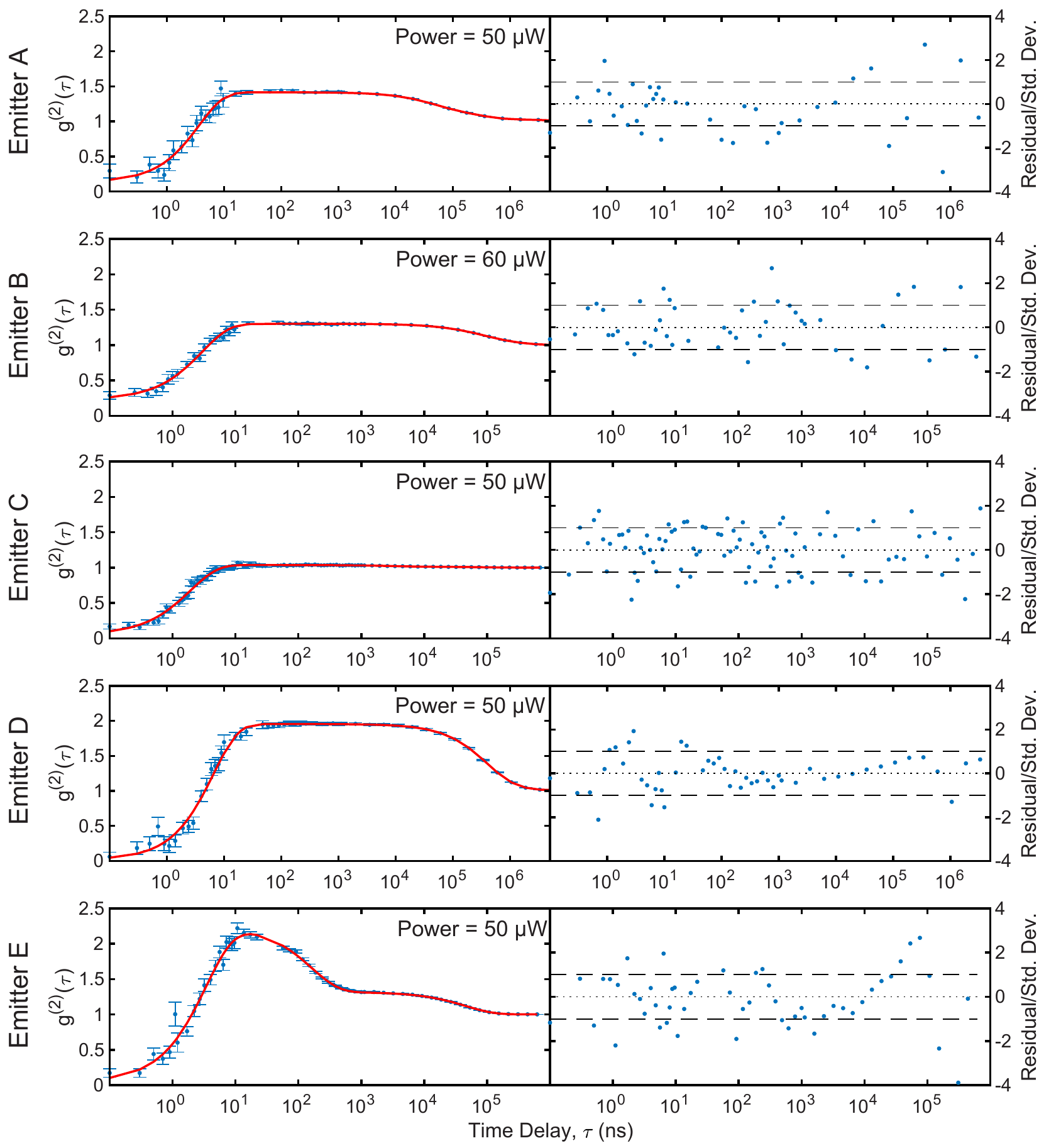}
  \caption{
  \textbf{Logarithmic time scale autocorrelation function.}
  \textbf{(Column 1)} Second-order autocorrelation function (blue points) calculated over a logarithmic time scale and fit using an empirical model discussed in the text (red curve).  Error bars represent Poissonian uncertainties based on the photon counts in each bin.
  \textbf{(Column 2)} Residual/Standard Deviation at each data point, obtained from the fit.}
  \label{fig:logScaleAutocorrelation}
\end{figure*}

Figure S\ref{fig:logScaleAutocorrelation} illustrates the second-order autocorrelation function calculated and analyzed over logarithmic time scale for each emitter and fit using an empirical model (as discussed in the main text), $\autocorr{\tau} = 1 - C_{1}e^{-|t|/\tau_{1}} + \sum_{i=2}^{n}C_{i}e^{-|t|/\tau_{i}}$.
The resultant fit parameters are summarized in Table~\ref{tab:logScaleParams}.

\begin{table*}[ht!] 
    \renewcommand{\thetable}{S\arabic{table}}
    \renewcommand\tablename{Table}
    \caption{\textbf{Logarithmic Time Scale Autocorrelation Function Fit Parameters.}}
    \label{tab:logScaleParams}
    \def\arraystretch{1.75}
    \begin{tabularx}{1\textwidth}
    { | X | X | X | X | X | X | }
    \rowcolor{lightGray}[\tabcolsep]
    \hline
    \textbf{Quantum} \par \textbf{Emitter} &
    \textbf{A} & \textbf{B} & \textbf{C} & \textbf{D} & \textbf{E} \\
    \hline
    \textbf{Excitation} \par \textbf{wavelength} &
    592 nm & 592 nm & 532 nm & 592 nm & 532 nm \\
    \hline
    $\mathbf{\chi^{2}_{red}}$ &
    1.801 & 1.202 & 1.085 & 1.031 & 2.872 \\ 
    \hline
    \textbf{n} &
    4 & 3 & 3 & 3 & 3 \\
    \hline
    $\mathbf{C_{1}}$ &
    1.285 $\pm$ 0.052 & 1.076 $\pm$ 0.029  & 0.988 $\pm$ 0.020 & 1.938 $\pm$ 0.039 & 2.204 $\pm$ 0.041 \\
    \hline
    $\mathbf{\tau_{1} (ns)}$ &
    3.66 $\pm$ 0.31 & 2.97 $\pm$ 0.16 & 1.91 $\pm$ 0.08 & 6.56 $\pm$ 0.34 & 3.50 $\pm$ 0.16 \\
    \hline
    $\mathbf{C_{2}}$ &
    0.208 $\pm$ 0.002 & 0.193 $\pm$ 0.001 & 0.023 $\pm$ 0.001 & 0.725 $\pm$ 0.028 & 0.928 $\pm$ 0.022 \\
    \hline
    $\mathbf{\tau_{2} (\micro s)}$ &
    41.200$\pm$ 0.617 & 75.088 $\pm$ 0.600 & 3.429 $\pm$ 0.218 & 354.67 $\pm$ 13.62 & 0.163 $\pm$ 0.005 \\
    \hline
    $\mathbf{C_{3}}$ &
    0.172 $\pm$ 0.002 & 0.108 $\pm$ 0.002 & 0.013 $\pm$ 0.000 & 0.219 $\pm$ 0.029 & 0.315 $\pm$ 0.002 \\
    \hline
    $\mathbf{\tau_{3} (\micro s)}$ &
    252.49 $\pm$ 2.64 & 264.32 $\pm$ 2.21 & 90.66 $\pm$ 2.27 & 972.30 $\pm$ 48.39 & 46.64 $\pm$ 0.31 \\
    \hline
    $\mathbf{C_{4}}$ &
    0.035 $\pm$ 0.001 & - & - & - & - \\
    \hline
    $\mathbf{\tau_{4} (ms)}$ &
    5.622 $\pm$ 0.317 & - & - & - & - \\
    \hline
    \end{tabularx}
\end{table*}

\subsubsection*{S5.2 Background Correction}
Background correction is done to account for the background and incoherent light detected along with the signal which can affect the autocorrelation function.
The following two background correction techniques are used:
\begin{enumerate}
    \item \textbf{Recording background from an offset spot}:
    The autocorrelation data was acquired from a background spot, same as the emitter. 
    The background spot is an offset spot, $\sim$1 $\micro$m from the emitter which seems to emulate the true background. 
    The background data was acquired for the equivalent time as the emitter, with all other experimental conditions such as excitation power kept same. 
    From the background data, the average background count rate was determined.
    This technique was applied to QEs A and B.
    
    \item \textbf{Recording background from a 2D scan}:
    Instead of recording background data of an offset spot, X and Y line ($\micro$-PL) scan of the emitter were acquired.
    Since the emitter is tracked during the acquisitions, the $\micro$-PL line scans along X and Y pass through the center of the emitter. 
    A 2D Gaussian fit to the line scans provides the background and signal of the spatially isolated emitter.
    This technique speeds up the data acquisition by a factor of 2 since the data from an offset spot is no longer needed to be acquired.
    This technique better approximates the background since the estimation is done right around the emitter, instead of an offset spot.
    This technique was applied to QEs C, D and E, and adopted as future autocorrelation background correction technique.
    \end{enumerate}

Using the background and signal acquired, following background correction equations from Ref. \cite{Brouri2000} are used to determine background-corrected autocorrelation function:
    
\begin{equation}
    \renewcommand{\theequation}{S\arabic{equation}}
    \rho = \frac{Signal}{Signal+Background}
\end{equation}
\begin{equation}
    \renewcommand{\theequation}{S\arabic{equation}}
    g^{(2)}_{bkgd}(\tau) = \frac{\autocorr{\tau}-(1-\rho^{2})}{\rho^{2}}
\end{equation}
where $\autocorr{\tau}$ is determined from the QE and $g^{(2)}_{bkgd}(\tau)$ is the background-corrected autocorrelation function.

\subsubsection*{S5.3 Akaike Information Criterion}

For a given set of data, the Akaike Information Criterion (AIC) is a quantitative method to determine the relative quality of a collection of statistical models.
Using the AIC, the relative quality of each of the models compared can be estimated.
It can be applied to any data and fitting routines.
Thus, it can be used to determine which model best fits to a given data set.
However, it does not determine absolute quality.
Using the AIC, the relative quality of Eq. 5 for different $n$ is determined.
The \textit{likelihood} of a model ($n=[2, 5]$) explaining the actual data is determined by comparing the AIC of each model.

The AIC is defined as:
\begin{equation}
    \renewcommand{\theequation}{S\arabic{equation}}
    \mathrm{AIC} = 2p - (\mathrm{log likelihood})
\end{equation}
where p is the number of fit parameters in the model. Here, Gauss log likelihood estimation is used.

The Gauss log likelihood is defined as:
\begin{equation}
    \renewcommand{\theequation}{S\arabic{equation}}
    -2\ln{(L_{G})} =
    \sum_{i}\biggl[ \frac{(c_{i}-m_{a}(x_{i}))^{2}}{\sigma_{i}^{2}}+ln(\sigma_{i}^{2})\biggl]+N \ln{(2\pi)}
\end{equation}
where $-2\ln{(L_{G})}$ is the Gauss log likelihood, $N$ is the number of data points, $c_{i}$ is the measured value of the $i^{th}$ data point, $m_{a}$ is the model function, $m_{a}(x_{i})$ is the model predicted value of the $i^{th}$ data point, $x_{i}$ is the residual of the $i^{th}$ data point and $\sigma_{i}$ is the standard deviation of the $i^{th}$ data point.Thus,
\begin{equation}
     \renewcommand{\theequation}{S\arabic{equation}}
    \mathrm{AIC} = 2p + 2\ln{(L_{G})}
\end{equation}
The likelihood of a model to explain the actual data from a collection of models is determined by calculating the weight of the model, defined as:
\begin{equation}
    \renewcommand{\theequation}{S\arabic{equation}}
    w_{i}=\exp{((\mathrm{AIC_{min}-AIC_{i})/2)}}
\end{equation}
where $w_{i}$ is the weight of the $i^{th}$ model, $\mathrm{AIC_{min}}$ is the minimum AIC value among all models and $\mathrm{AIC_{i}}$ is the AIC value of the $i^{th}$ model.

The model with highest $w$ best explains the actual data, amongst the collection of models.
However, for relatively close values of two $w_{i}$, a simpler model could be selected.
Thus, the AIC is used to determine relative quality of $n$ for $n=[2, 5]$.

\subsection*{S6. Lifetime Measurement (QE A)}

\begin{figure*}[ht!]
  \renewcommand\figurename{Figure S}
  \let\nobreakspace\relax
  \centering
  \includegraphics[width=3.375in]{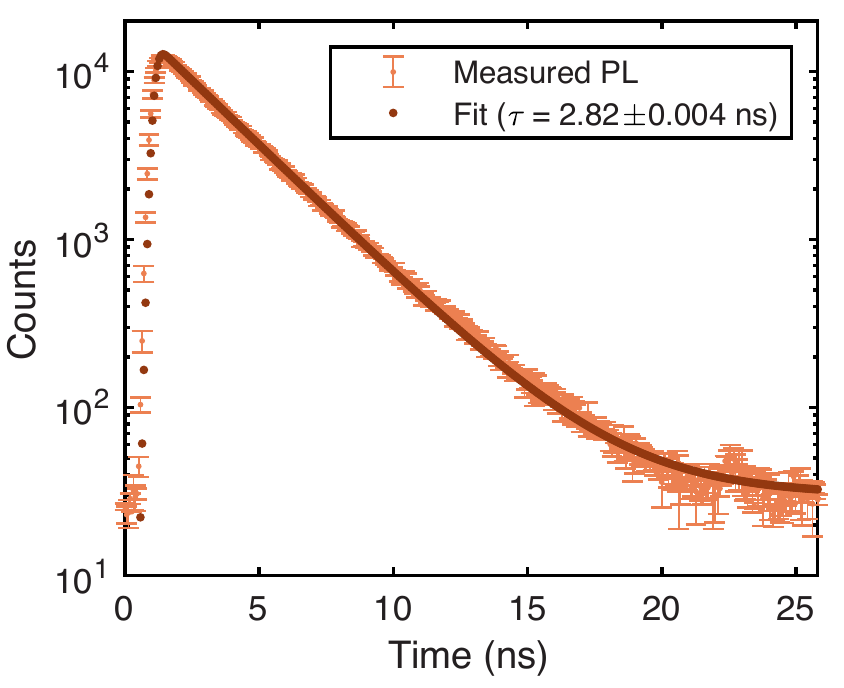}
  \caption{\textbf{Quantum emitter A: lifetime measurement.}
  Time-resolved PL (light orange circles) and fit (dark orange circle) to the data. 
  The error bars represent one standard deviation.
  }
  \label{fig:lifetime}
\end{figure*}

Figure S\ref{fig:lifetime} shows time-resolved PL of the QE A.
Time-resolved PL is acquired by using a pulsed laser (NKT Photonics, Fianium Whitelase) with the excitation wavelength centered $\sim$580 nm and 40 MHz pulse rate.
The PL is recorded in histogram mode with the TCSPC module (PicoQuant, PicoHarp 300).
The lifetime is obtained by fitting the convolution of the IRF and an empirical model ($A \times \exp{(-t/\tau)}+B$) to the data, given as $\mathrm{IRF} * (A \times \exp{(-t/\tau)}+B)$.
The IRF is obtained in same configuration as the time-resolved PL of the QE.
The lifetime is measured as 2.82$\pm$0.004 ns (355 MHz).

\subsection*{S7. Photon Emission Statistics}\label{sec:SI_PhotonEmissionStats}

\subsubsection*{S7.1 Nitrogen Vacancy Center in Diamond}

\begin{figure*}[ht]
  \renewcommand\figurename{Figure S}
  \let\nobreakspace\relax
  \centering
  \includegraphics[width=3.375in]{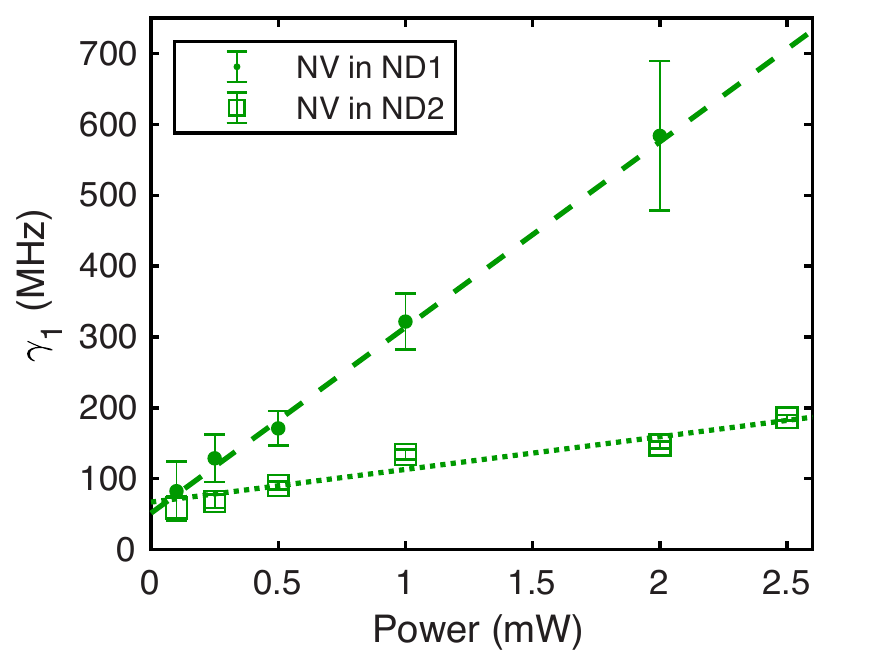}
  \caption{\textbf{Antibunching rate of NV centers in nanodiamonds.}
  The antibunching rate (denoted by circles and squares) is measured as a function of excitation power.
  The lines (dashed and dotted) are fits to the rates.
  The error bars represent one standard deviation.
  }
  \label{fig:ndExample}
\end{figure*}

The antibunching rate as a function of excitation power is typically linear, where the y-intercept provides the inverse of the radiative lifetime and the slope relates the pumping power to excitation rate.
This is the case for the NV center in diamond. 
The autocorrelation function of single NVs in multiple nanodiamonds is calculated to verify linear dependence of antibunching rate on excitation power.
Figure S\ref{fig:ndExample} shows the antibunching rate, $\gamma_{1}$ as a function of excitation power of NV center in two nanodiamonds.
The nanodiamonds were dropcasted on a silicon wafer and probed for single NVs.
The nanodiamond sample was studied in the same setup as the h-BN samples discussed in the main text.
The data acquisition and analysis is as discussed in the main text (no IRF correction was performed).
The data is acquired using green (532 nm) excitation.
To check for linear dependence, the rates are fit using Eq. 6a.
The y-intercept is calculated to be 51 MHz (NV in ND1) and 67 MHz (NV in ND2), which agrees with that reported in the literature \cite{Berthel2015}.

\subsubsection*{S7.2 QE A: $\gamma_{4}$ and $C_{4}$}

\begin{figure*}[ht]
  \renewcommand\figurename{Figure S}
  \let\nobreakspace\relax
  \centering
  \includegraphics[width=3.375in]{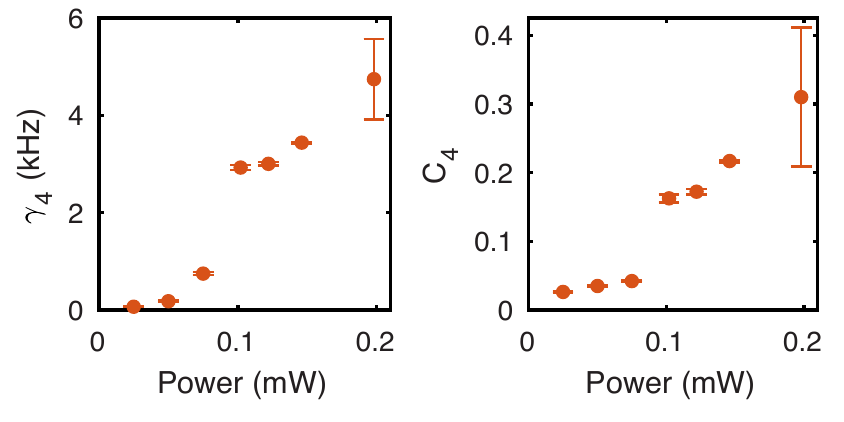}
  \caption{\textbf{Quantum emitter A: $\gamma_{4}$ and $C_{4}$ for orange excitation.}
  The error bars denote one standard deviation.
  }
  \label{fig:emAtau4}
\end{figure*}

Figure S\ref{fig:emAtau4} shows bunching rate $\gamma_{4}$ and amplitude $C_{4}$ for orange excitation.
For orange excitation, QE A is modeled by 4 timescales ($n=4$).

\begin{figure*}[ht!]
  \renewcommand\figurename{Figure S}
  \let\nobreakspace\relax
  \centering
  \includegraphics[width=3.375in]{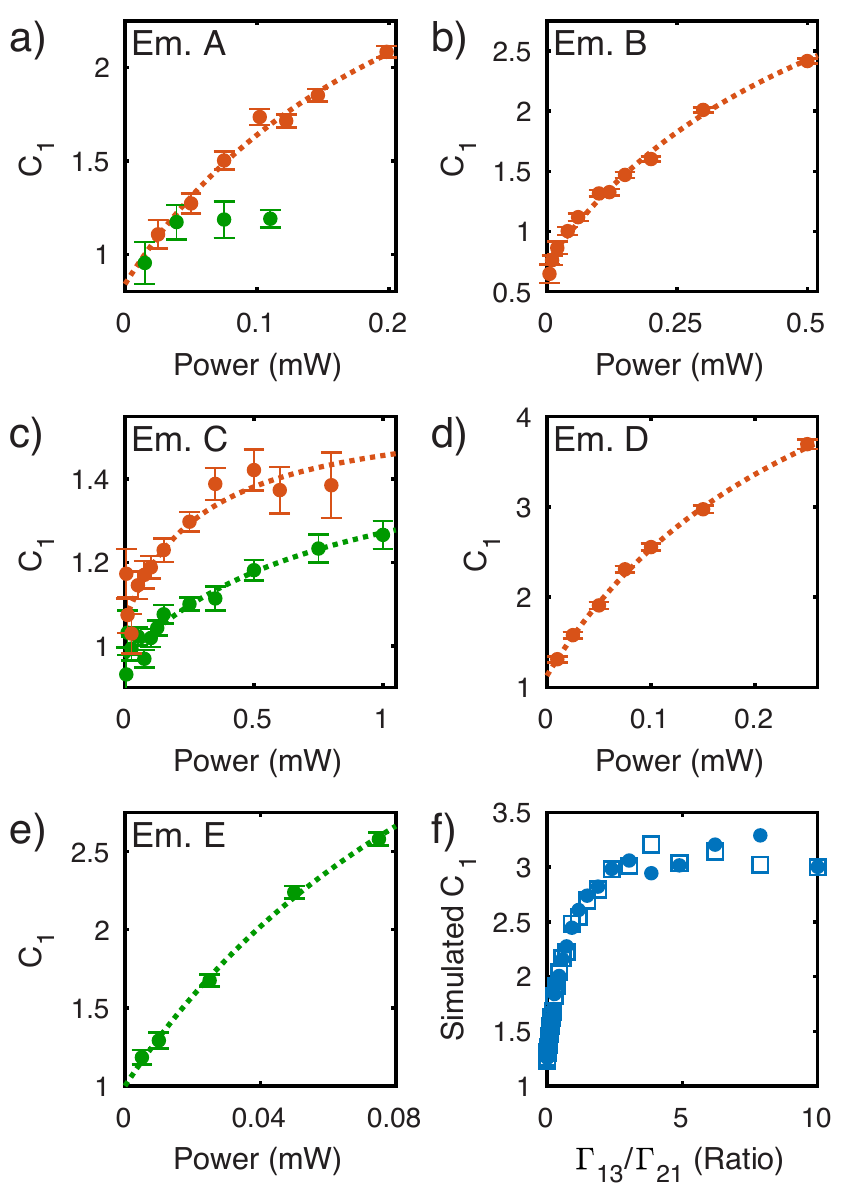}
  \caption{\textbf{Antibunching amplitudes.}
  \textbf{(a-e)} Antibunching amplitudes (circles) of QEs A to E.
  The error bars represent one standard deviation.
  The dashed lines are fits using an empirical model discussed in the text.
  \textbf{(f)} Simulated antibunching amplitudes as a function of excitation rate for spontaneous (circles) and pumped (squares) transition mechanism discussed in the main text.
  }
  \label{fig:antibunchingAmplitudes}
\end{figure*}

\subsubsection*{S7.3 Antibunching Amplitude}

Figure S\ref{fig:antibunchingAmplitudes}(a-e) presents the excitation power dependence of the antibunching amplitude, $C_{1}$ of the QEs.
Figure S\ref{fig:antibunchingAmplitudes}(f) presents the simulated antibunching amplitude for spontaneous and pumped transition mechanisms.
The dashed lines are fits using the first order saturation model (Eq. 6b).
The fit results are shown in Table~\ref{tab:empiricalFits2}.

\subsubsection*{S7.4: Simulations}

\begin{figure*}[h!]
  \renewcommand\figurename{Figure S}
  \let\nobreakspace\relax
  \centering
  \includegraphics[width=7in]{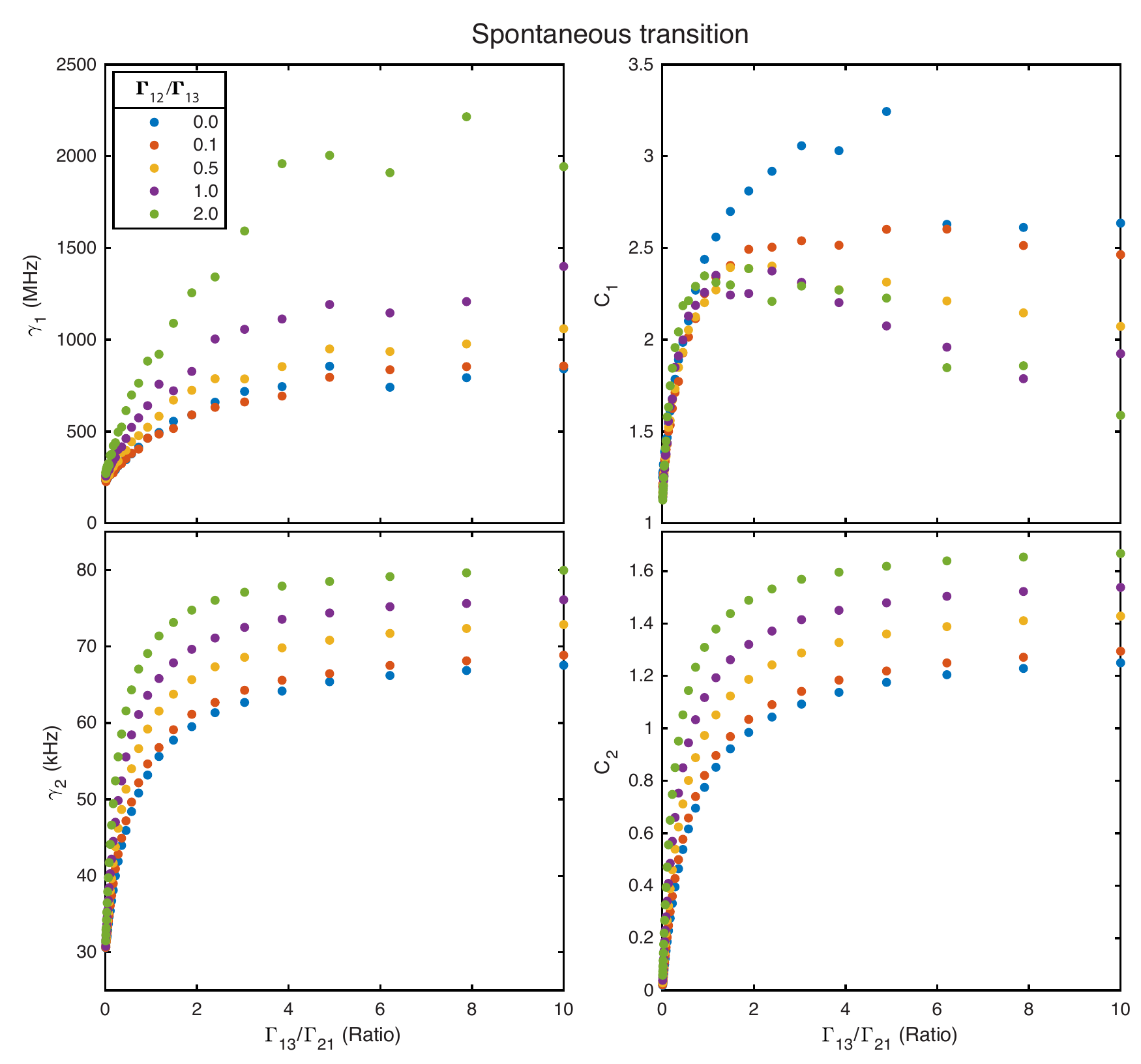}
  \caption{\textbf{Simulation of optical dynamics for different $\Gamma_{12}/\Gamma_{13}$.}
  Antibunching and bunching parameters resulting from simulation of the model discussed in main text for $\Gamma_{12}$ as a factor of $\Gamma_{13}$ for spontaneous transition.
  Simultaneous excitation to level 2 and level 3 takes place at different rates which is ratio of the two rates.
  }
  \label{fig:simulationRatio}
\end{figure*}

Figure S\ref{fig:simulationRatio} presents result of simulations for $\Gamma_{12}/\Gamma_{13} = [0,2]$, for the case of spontaneous transition via level 4.
As highlighted in the main text, the result of simulations for different $\Gamma_{12}/\Gamma_{13}$ are qualitatively similar.

\subsubsection*{S7.5: Simulating optical dynamics of QE A}

\begin{figure*}[htb!]
  \renewcommand\figurename{Figure S}
  \let\nobreakspace\relax
  \centering
  \includegraphics[width=7in]{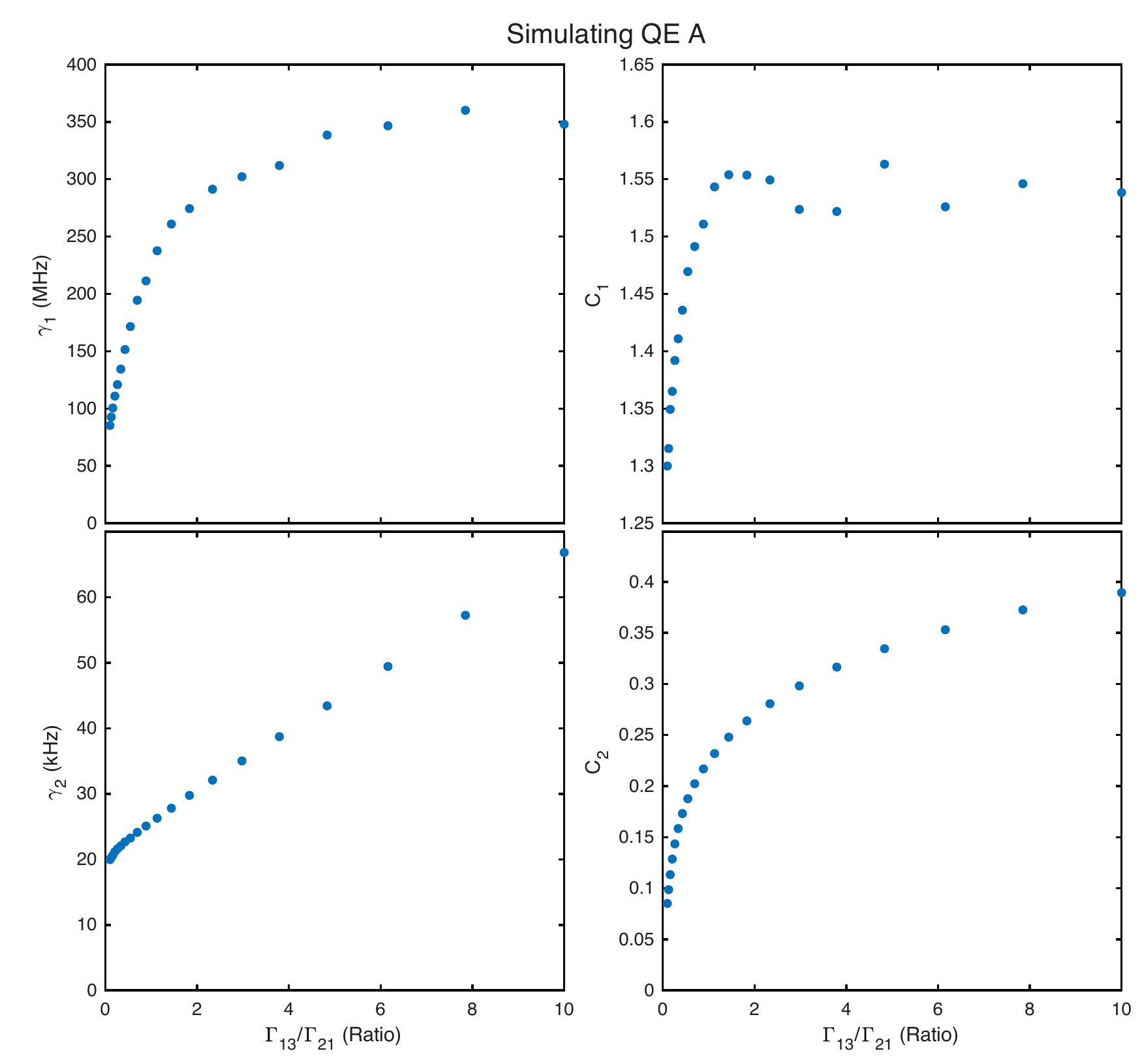}
  \caption{\textbf{Simulating optical dynamics of QE A.}
  Antibunching and bunching parameters resulting from simulation of the model discussed in main text by setting the radiative and nonradiative rates to best recreate the observed optical dynamics of QE A.
  The simulation is using the parameters discussed in the text: $\Gamma_{21}=300$ MHz, $\Gamma_{32}=60$ MHz, $\Gamma_{12}=0$ MHz, $\kappa_{24,0}=24$ kHz, $\beta_{24}=9$ kHz, $\kappa_{41,0}=18$ kHz and $\beta_{41}=3$ kHz.
  The results are plotted as a function of $\Gamma_{13}/\Gamma_{21}$, where $\Gamma_{21}$is a fixed parameter.
  }
  \label{fig:simulating_QEA}
\end{figure*}

Figure S\ref{fig:simulating_QEA} shows the result of simulations to recreate the observed optical dynamics of QE A.
The same model discussed in the main text is used in the simulations (Fig. 5(a)).
The radiative and nonradiative rates are chosen such that the resultant dynamics best recreates the observed optical dynamics of QE A (1$^{st}$ column of Fig. 4).
There are two important qualitative features of simulating QE A that differ from the simulations presented in Fig. 5 of the main text.  
The first is that when $\kappa_{32}<\Gamma_{21}$, we find that the observed antibunching rate, $\gamma_1$, is less than the spontaneous emission rate, $\Gamma_{21}$, over a wide range of power settings. 
Here, $\Gamma_{21}=300$~MHz and $\kappa_{32}=60$~MHz.
The second important feature of simulating QE A is that the nonradiative transition mechanism involves both the spontaneous and optically pumped components.
The nonradiative rates $\kappa_{24}$ and $\kappa_{41}$ are given the following form:
\begin{equation}
    \renewcommand{\theequation}{S\arabic{equation}}
    \kappa_{ij} = \kappa_{ij,0} + \beta_{ij}\frac{\Gamma_{13}}{\Gamma_{21}}
\end{equation}
where $\kappa_{ij,0}$ is the spontaneous emission rate, and $\beta_{ij}$ is a scaling factor for the optically pumped transition, giving the corresponding transition rate at saturation when $\Gamma_{13}=\Gamma_{21}$.
Here, we set $\kappa_{24,0}=24$~kHz, $\kappa_{41,0}=18$~kHz, $\beta_{24}=9$~kHz, and $\beta_{41}=3$~kHz.
The combination of spontaneous and pumped transition quantiatively reproduces the non-zero offset of the bunching rate and the quasi-linear power scaling.

\clearpage
\subsection*{S8: General state transition diagram}

\begin{figure*}[h]
  \renewcommand\figurename{Figure S}
  \let\nobreakspace\relax
  \centering
  \includegraphics[width=3in]{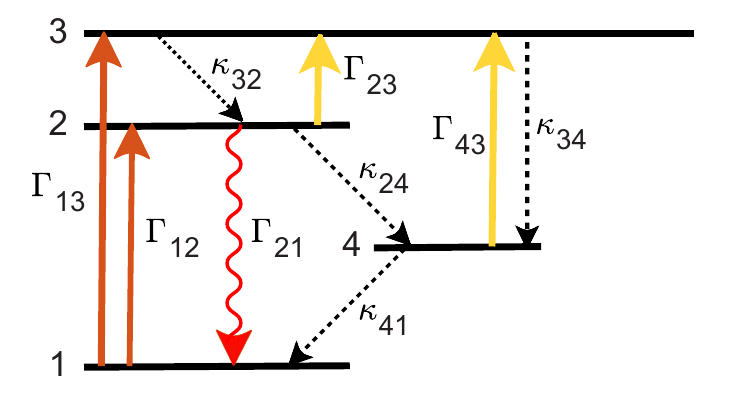}
  \caption{\textbf{General electronic level structure.}
  }
  \label{fig:extendedLevelStructure}
\end{figure*}

Figure S\ref{fig:extendedLevelStructure} generalizes the electronic level structure and transition diagram shown in the main text (Fig. 5(a)).
The general version shows how effective transitions that are proportional to optical pumping power (e.g., optically pumped transitions $\kappa_{24}$ and $\kappa_{41}$) can result from combinations of other possible transitions such as re-pumping from level $2\rightarrow3$ with rate $\Gamma_{23}$, re-pumping from level $4\rightarrow3$ with rate $\Gamma_{43}$ and nonradiative transition from level $3\rightarrow4$ with rate $\kappa_{34}$. 

\subsection*{S9. Calculation of the capture coefficient for the boron dangling bond}

We calculate the nonradiative capture coefficient $C_n$ for the capture of an electron from the conduction band into the boron dangling bond using the formalism implemented in the Nonrad code~\cite{Turiansky2021b}.
We will focus on the ground state [level 1 in Fig. 5(a)] and the optically active excited state [level 2 in Fig. 5(a)] of the dangling bond, which are separated by 2.06~eV~\cite{Turiansky2019a}.
In equilibrium, the dangling bond is in the negative charge state and is occupied by two electrons.
When the excitation energy is sufficiently large, an electron can be excited into the conduction band and the dangling bond is photoionized, changing the charge state from negative to neutral [process $\Gamma_{13}$ in Fig. 5(a)].
Subsequent re-capture of this electron returns the dangling bond to the negative charge state.
We consider two potential scenarios for this nonradiative process mediated by electron-phonon coupling:
(1) The electron is captured directly into the ground state of the dangling bond, with rate $\kappa_{31}$ [not depicted in Fig. 5(a)], or (2) the electron is captured into the excited state of the dangling bond [$\kappa_{32}$ in Fig. 5(a)].
Process (2) puts the defect in the optically active excited state, from which a photon can then be emitted, with an emissive dipole unaligned with the absorptive dipole.

To evaluate the nonradiative capture coefficients, we extract several parameters from our density-functional theory calculations:
the transition energy, the mass-weighted root-mean-square difference in atomic geometries $\Delta Q$, the phonon frequencies in the initial ($i$) and final ($f$) states $\Omega_{i/f}$, and the electron-phonon coupling matrix element $W_{if}$.
The transition level for capture into excited state, which is used to determine the transition energy, is above the conduction-band minimum, while the single-particle states are in the gap.
For the purposes of our capture coefficient evaluation, we shift the transition energy to be consistent with the 200~meV difference observed experimentally~\cite{Jungwirth2017};
we verified that the conclusions are insensitive to the choice of the energy shift.
The degeneracy factor in the nonradiative rate~\cite{Turiansky2021b} is set to 1 since the dangling bond does not possess any configurational degeneracy.
A scaling factor that accounts for charged defect interactions [see Sec.~III.~E. of Ref.~\cite{Alkauskas2014}] is not necessary in this case because capture occurs in the neutral charge state and the electron-phonon coupling is evaluated in the neutral charge state.

At room temperature, we calculate $C_n$ for capture into the ground state to be $1.2 \times 10^{-12}$~cm$^3$~s$^{-1}$ and into the excited state to be $1.2 \times 10^{-7}$~cm$^3$~s$^{-1}$.
We can thus safely assume that capture into the excited state will dominate.
These capture coefficients are larger than typical radiative capture coefficients, which are on the order of $10^{-13}$ - $10^{-14}$~cm$^3$~s$^{-1}$~\cite{Dreyer2020}, justifying our implicit assumption of nonradiative rather than radiative capture.

\begin{figure*}[htb!]
  \renewcommand\figurename{Figure S}
  \let\nobreakspace\relax
  \centering
  \includegraphics[width=7in]{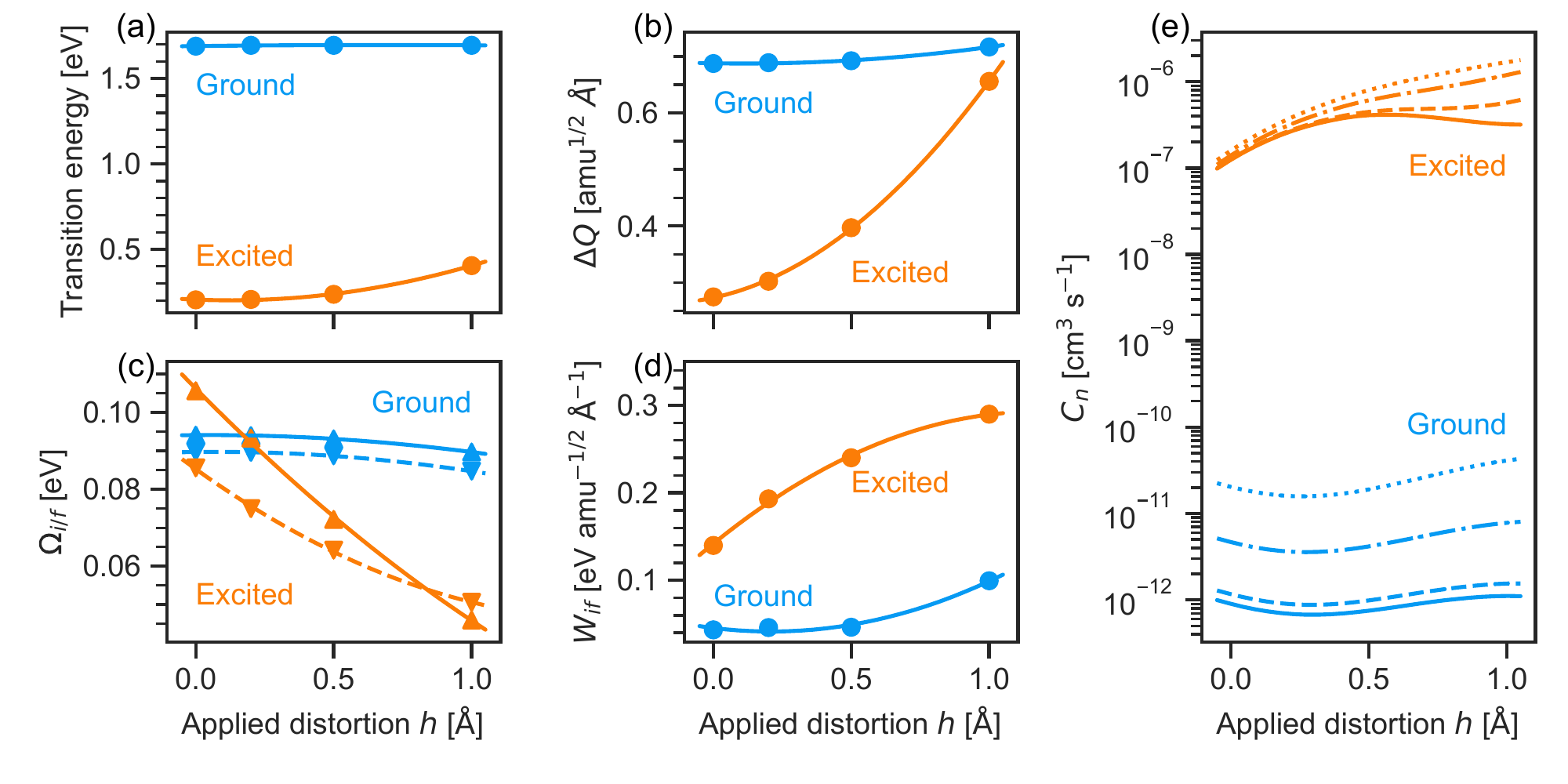}
  \caption{\textbf{Calculated parameters and capture coefficient for capture of an electron from the conduction band into the neutral boron dangling bond, as a function of applied distortion $h$.}
    Calculated \textbf{(a)} transition energy, \textbf{(b)} mass-weighted root-mean-square difference in atomic geometries, \textbf{(c)} initial (up triangle) and final (down triangle) phonon frequencies, and \textbf{(d)} electron-phonon coupling matrix elements.
    The lines are a quadratic fit to the calculated parameters and are intended to guide the eye.
    The calculated \textbf{(e)} electron capture coefficient at 10~K (solid), 300~K (dashed), 600~K (dashed dotted), and 900~K (dotted).
    The parameters for capture into the ground state are shown in blue and the excited state are shown in orange.
  }
  \label{fig:capCoeff}
\end{figure*}

Previous work has already demonstrated that out-of-plane distortions are important for understanding both the symmetry~\cite{Turiansky2019a} and transition rates~\cite{Turiansky2021} of the boron dangling bond.
Here we include the effect of out-of-plane distortions on the capture coefficient, following the methodology of Ref.~\cite{Turiansky2021}.
In short, a plane neighboring the dangling bond is bent to create a ``bubble'', with the height of the bubble being referred to as the applied distortion $h$.
The bubble induces an out-of-plane distortion in the dangling bond and allows us to study its effect on the capture coefficient.
The influence of the out-of-plane distortion on the calculated parameters is shown in Fig. S\ref{fig:capCoeff}.
For comparison purposes, we will use the average at room temperature $4 \times 10^{-7}$ cm$^3$ s$^{-1}$ as a representative value for capture into the excited state.

\clearpage
\subsection*{S10. Empirical Fits to Photon Emission Statistics}

Table~\ref{tab:empiricalFits1} and \ref{tab:empiricalFits2} summarizes the results of fits in Fig. 4 and Fig. S\ref{fig:antibunchingAmplitudes}.


\begin{table*}[htb!]
    \renewcommand{\thetable}{S\arabic{table}}
    \renewcommand\tablename{Table}
    \caption{\textbf{Results of Fitting Empirical Functions in Eq. 6a - 6d to PECS: Rates}}
    \label{tab:empiricalFits1}
    \def\arraystretch{1.75}
    \begin{scriptsize}
    \begin{tabularx}{1\textwidth}{ | p{0.5cm} | X | X | X | X | X | }
    \rowcolor{lightGray}[\tabcolsep]
    
    \hline
    \textbf{QE} &
    \textbf{A} & \textbf{B} & \textbf{C} & \textbf{D} & \textbf{E} \\
    
    \hline
    
    $\mathbf{\gamma_{1}}$ &
    
    532 nm: \par \underline{1st Order Saturation} \par
    $R_{0}=0.00\pm261.5$ MHz \par $R_{sat}=426.9\pm230.6$ MHz \par $P_{sat}=17.23\pm23.90$ $\micro$W \par
    \par \vspace{1.5em} 
    592 nm: \par \underline{Saturation} \par $R_{0}=0.00\pm167.4$ MHz \par $R_{sat}=411.4\pm154.1$ MHz \par $P_{sat}=26.32\pm20.10$ $\micro$W \par &
    
    592 nm: \par \underline{1st Order Saturation} \par $R_{0}=276.6\pm1.0$ MHz \par $R_{sat}=156.5\pm2.19$ MHz \par $P_{sat}=135.8\pm7.23$ MHz$/\micro$W &
    
    532 nm: \par \underline{2nd Order Saturation} \vspace{0.0em} \par $R_{0}=452.0\pm47.1$ MHz \par $m_{0}=3.06\pm3.32$ MHz$/\micro$W \par $m_{1}=0.3\pm0.07$ MHz$/\micro$W$^2$ \par $P_{sat}=52.2\pm60.8$ $\micro$W
    \par \vspace{1.5em} 
    592 nm: \par \underline{2nd Order Saturation} \vspace{0.0em} \par $R_{0}=502.3\pm46.8$ MHz \par $m_{0}=4.48\pm2.84$ MHz$/\micro$W \par $m_{1}=0.1\pm0.1$ MHz$/\micro$W$^2$ \par $P_{sat}=75.3\pm57.2$ $\micro$W &
    
    592 nm: \par \underline{Linear} \par $R_{0}=135.8\pm4.7$ MHz \par $m_{0}=0.18\pm0.03$ MHz$/\micro$W &
    
    532 nm: \par \underline{Linear} \par $R_{0}=227.7\pm12.5$ MHz \par $m_{0}=1.0\pm0.24$ MHz$/\micro$W \\
    
    \hline
    
    $\mathbf{\gamma_{2}}$ &
    
    592 nm: \par \underline{Linear} \par $R_{0}=16.7\pm0.4$ kHz \par $m_{0}=0.15\pm0.005$ kHz/$\micro$W &
    
    592 nm: \par \underline{2nd Order Saturation} \par $R_{0}=0.0\pm9.87$ kHz \par $m_{0}=3.13\pm7.02$ kHz/$\micro$W \par $m_{1}=0.0363\pm0.0028$ kHz/$\micro$W$^2$ \par $P_{sat}=3.66\pm5.29$ $\micro$W &
     
    532 nm: \par \underline{1st Order Saturation} \par $R_{0}=0.0\pm4.3$ MHz \par $R_{sat}=24.1\pm4.3$ MHz \par $P_{sat}=119.4\pm40.8$ $\micro$W
    \par \vspace{1.5em} 
    592 nm: \par \underline{1st Order Saturation} \par $R_{0}=0.0\pm0.8$ MHz \par $R_{sat}=328.5\pm4.2$ MHz \par $P_{sat}=1177.2\pm32.7$ $\micro$W &
    
    592 nm: \par \underline{Linear} \par $R_{0}=1.74\pm0.08$ kHz \par $m_{0}=0.02\pm0.001$ kHz/$\micro$W &
    
    532 nm: \par \underline{Linear} \par $R_{0}=3.07\pm0.0.2$ MHz \par $m_{0}=0.06\pm0.004$ kHz/$\micro$W \\
    
    \hline
    
    $\mathbf{\gamma_{3}}$ &
    
    592 nm: \par \underline{Quadratic} \par $R_{0}=1.4\pm1.4$ kHz \par $m_{0}=0.00\pm0.03$ kHz$/\micro$W \par $m_{1}=0.0007\pm0.0001$ kHz$/\micro$W$^2$ & 
    
    592 nm: \par \underline{2nd Order Saturation} \par $R_{0}=0.0\pm0.63$ kHz \par $m_{0}=1.10\pm0.43$ kHz$/\micro$W \par $m_{1}=0.0076\pm0.0001$ kHz$/\micro$W$^2$ \par $P_{sat}=3.01\pm0.66$ $\micro$W  & 
     
    532 nm: \par \underline{1st Order Saturation} \par $R_{0}=0.0\pm0.6$ kHz \par $R_{sat}=4336.2\pm2.8$ kHz \par $P_{sat}=540.0\pm0.8$ $\micro$W
    \par \vspace{1.5em} 
    592 nm: \par \underline{1st Order Saturation} \par $R_{0}=0.0\pm0.2$ kHz \par $R_{sat}=2815.4\pm61.8$ kHz \par $P_{sat}=1550.4\pm42.2$ $\micro$W & 
    
    592 nm: \par \underline{Linear} \par $R_{0}=0.82\pm0.03$ kHz \par $m_{0}=0.002\pm0.000$ kHz$/\micro$W &
    
    532 nm: \par \underline{Quadratic} \vspace{0.0em} \par $R_{0}=0.00\pm1.05$ kHz \par $m_{0}=0.211\pm0.075$ kHz$/\micro$W \par $m_{1}=0.004\pm0.0009$ kHz$/\micro$W$^2$ \\    
    \hline
    \end{tabularx}
    \end{scriptsize}
\end{table*}

\begin{table*}[ht]
    \renewcommand{\thetable}{S\arabic{table}}
    \renewcommand\tablename{Table}
    \caption{\textbf{Results of Fitting Empirical Functions in Eq. 6a - 6d to PECS: Amplitudes}}
    \label{tab:empiricalFits2}
    \def\arraystretch{1.75}
    \begin{scriptsize}
    \begin{tabularx}{1\textwidth}{ | p{0.5cm} | X | X | X | X | X | }
    \rowcolor{lightGray}[\tabcolsep]
    
    \hline
    \textbf{QE} &
    \textbf{A} & \textbf{B} & \textbf{C} & \textbf{D} & \textbf{E} \\

    \hline
    
    $\mathbf{C_{1}}$ &
    
    592 nm: \par \underline{1st Order Saturation} \par $R_{0}=0.84\pm0.13$ \par $R_{sat}=2.74\pm0.6$ \par $P_{sat}=241.95\pm128.4$ $\micro$W & 
    
    592 nm: \par \underline{1st Order Saturation} \par $R_{0}=0.74\pm0.03$ \par $R_{sat}=3.75\pm0.25$ \par $P_{sat}=617.5\pm78.5$ $\micro$W &
    
    532 nm: \par \underline{1st Order Saturation} \par $R_{0}=0.97\pm0.02$ \par $R_{sat}=0.55\pm0.16$ \par $P_{sat}=841.0\pm483.5$ $\micro$W  
    \par \vspace{1.5em} 
    592 nm: \par \underline{1st Order Saturation} \par $R_{0}=1.06\pm0.03$ \par $R_{sat}=0.53\pm0.1$ \par $P_{sat}=304.4\pm148.5$ $\micro$W & 
    
    592 nm: \par \underline{1st Order Saturation} \par $R_{0}=1.13\pm0.04$ \par $R_{sat}=5.45\pm0.47$ \par $P_{sat}=284.84\pm44.06$ $\micro$W &
    
    532 nm: \par \underline{1st Order Saturation} \par $R_{0}=1.0\pm0.06$ \par $R_{sat}=4.49\pm1.11$ \par $P_{sat}=136.17\pm54.55$ $\micro$W \\
    
    \hline
    
    $\mathbf{C_{2}}$ &
    
    592 nm: \par \underline{1st Order Saturation} \par $R_{0}=0.0\pm1.82$ \par $R_{sat}=0.95\pm4.15$ \par $P_{sat}=137.4\pm1745.8$ $\micro$W &
     
    592 nm: \par \underline{Linear} \par $R_{0}=0.029\pm0.416$ \par $m_{0}=0.0028\pm0.0021$ $\micro$W$^{-1}$ &
     
    532 nm: \par \underline{Linear}  \par $R_{0}=0.000\pm0.004$ \par $m_{0}=0.00016\pm0.00001$ $\micro$W$^{-1}$  
    \par \vspace{1.5em} 
    592 nm: \par \underline{1st Order Saturation} \par $R_{0}=0.0\pm0.84$ \par $R_{sat}=0.32\pm2.82$ \par $P_{sat}=799.3\pm18276.6$ $\micro$W &
    
    592 nm: \par \underline{1st Order Saturation} \par $R_{0}=0.0\pm0.02$ \par $R_{sat}=5.59\pm0.29$ \par $P_{sat}=321.0\pm27.8$ $\micro$W &
    
    532 nm: \par \underline{Linear} \par $R_{0}=0.065\pm0.012$ \par $m_{0}=0.017\pm0.000$ $\micro$W$^{-1}$ \\ 
    
    \hline
    
    $\mathbf{C_{3}}$ &
    
    592 nm: \par \underline{1st Order Saturation} \par $R_{0}=0.0\pm0.017$ \par $R_{sat}=0.252\pm0.015$ \par $P_{sat}=21.52\pm3.06$ $\micro$W &

    592 nm: \par \underline{1st Order Saturation} \par $R_{0}=0.0\pm0.0013$ \par $R_{sat}=0.259\pm0.002$ \par $P_{sat}=81.53\pm2.25$ $\micro$W &
    
    532 nm: \par \underline{1st Order Saturation} \par $R_{0}=0.004\pm0.574$ \par $R_{sat}=0.27\pm3.34$ \par $P_{sat}=581.9\pm17034.4$ $\micro$W \par 
    \par \vspace{1.5em} 
    592 nm: \par \underline{1st Order Saturation} \par $R_{0}=0.004\pm0.0004$ \par $R_{sat}=0.531\pm0.002$ \par $P_{sat}=163.6\pm2.3$ $\micro$W &
    
    592 nm: \par \underline{1st Order Saturation} \par $R_{0}=0.0\pm0.35$ \par $R_{sat}=0.20\pm0.35$ \par $P_{sat}=3.29\pm7.65$ $\micro$W & 
    
    532 nm: \par \underline{1st Order Saturation} \par $R_{0}=0.0042\pm0.0006$ \par $R_{sat}=1.85\pm0.1$ \par $P_{sat}=251.56\pm16.45$ $\micro$W \\    
    \hline
    \end{tabularx}
    \end{scriptsize}
\end{table*}

\newpage
\bibliography{arxiv_si}